\def\mearth{{\rm\,M_\oplus}}
\def\rearth{{\rm\,R_\oplus}}
\newcommand{\scale}[3]{\left(\frac{#1}{#2}\right)^{#3}}

\documentclass[preprint2]{proto}
\usepackage{times}
\usepackage{amsmath}

\begin{document}

\title{\textbf{\LARGE TERRESTRIAL PLANET FORMATION AT HOME AND ABROAD}}

\author {\textbf{\large Sean N. Raymond}}
\affil{\small\em Laboratoire d'Astrophysique de Bordeaux,
CNRS and Universit{\'e} de Bordeaux,
33270 Floirac, France}

\author {\textbf{\large Eiichiro Kokubo}}
\affil{\small\em Division of Theoretical Astronomy,
National Astronomical Observatory of Japan,
Osawa, Mitaka, Tokyo, 181-8588, Japan}

\author {\textbf{\large Alessandro Morbidelli}}
\affil{\small\em Laboratoire Lagrange, Observatoire de la Cote d'Azur, Nice, France}

\author {\textbf{\large Ryuji Morishima}}
\affil{\small\em University of California, Los Angeles,
Institute of Geophysics and Planetary Physics,
Los Angeles, CA 90095, USA}

\author {\textbf{\large Kevin J. Walsh}}
\affil{\small\em Southwest Research Institute, Boulder, Colorado 80302, USA}

\begin{abstract}
\baselineskip = 11pt
\leftskip = 0.65in 
\rightskip = 0.65in
\parindent=1pc
{\small We review the state of the field of terrestrial planet formation with the goal of understanding the formation of the inner Solar System and low-mass exoplanets.  We review the dynamics and timescales of accretion from planetesimals to planetary embryos and from embryos to terrestrial planets.  We discuss radial mixing and water delivery, planetary spins and the importance of parameters regarding the disk and embryo properties.  Next, we connect accretion models to exoplanets.  We first explain why the observed hot Super Earths probably formed by in situ accretion or inward migration.  We show how terrestrial planet formation is altered in systems with gas giants by the mechanisms of giant planet migration and dynamical instabilities.  Standard models of terrestrial accretion fail to reproduce the inner Solar System.  The ``Grand Tack'' model solves this problem using ideas first developed to explain the giant exoplanets.  Finally, we discuss whether most terrestrial planet systems form in the same way as ours, and highlight the key ingredients missing in the current generation of simulations.  
 \\~\\~\\~}
\end{abstract}

\section{\textbf{INTRODUCTION}}

The term ``terrestrial planet'' evokes landscapes of a rocky planet like Earth or Mars but given recent discoveries it has become somewhat ambiguous.  Does a $5 \mearth$ Super Earth count as a terrestrial planet?   What about the Mars-sized moon of a giant planet?  These objects are terrestrial planet-sized but their compositions and corresponding landscapes probably differ significantly from our terrestrial planets'.  In addition, while Earth is thought to have formed via successive collisions of planetesimals and planetary embryos, the other objects may have formed via different mechanisms.  For instance, under some conditions a $10 \mearth$ or larger body can form by accreting only planetesimals, or even only cm-sized pebbles.  In the context of the classical stages of accretion this might be considered a ``giant embryo'' rather than a planet (see \S 7.1).

What criteria should be used to classify a planet as terrestrial?  A bulk density higher than a few $g\,cm^{-3}$ probably indicates a rock-dominated planet, but densities of low-mass exoplanets are extremely challenging to pin down~\citep[see][]{marcy13}.  A planet with a bulk density of $0.5-2 g\,cm^{-3}$ could either be rocky with a small H-rich envelope or an ocean planet~\citep{fortney07,valencia07,adams08}.  Bulk densities larger than $3 g\,cm^{-3}$ have been measured for planets as massive as $10-20 \mearth$, although higher-density planets are generally smaller~\citep{weiss13}.  Planets with radii $R \lesssim 1.5-2 \rearth$ or masses $M \lesssim 5-10 \mearth$ are likely to preferentially have densities of $3 g\,cm^{-3}$ or larger and thus be rocky~\citep{weiss13b,lopez13}.

In this review we address the formation of planets in orbit around stars that are between roughly a lunar mass ($\sim 0.01 \mearth$) and ten Earth masses.  Although the compositions of planets in this mass range certainly vary substantially, these planets are capable of having solid surfaces, whether they are covered by thick atmospheres or not.  These planets are also below the expected threshold for giant planet cores~\citep[e.g.][]{lissauer07}.  We refer to these as terrestrial planets.  We start our discussion of terrestrial planet formation when planetesimals have already formed; for a discussion of planetesimal formation please see the chapter by Johansen et al.  

Our understanding of terrestrial planet formation has undergone a dramatic improvement in recent years.  This was driven mainly by two factors: increased computational power and observations of extra-solar planets.  Computing power is the currency of numerical simulations, which continually increase in resolution and have become more and more complex and realistic.  At the same time, dramatic advances in exoplanetary science have encouraged many talented young scientists to join the ranks of the planet formation community.  This manpower and computing power provided a timely kick in the proverbial butt.

Despite the encouraging prognosis, planet formation models lag behind observations.  Half of all Sun-like stars are orbited by close-in ``super Earths'', yet we do not know how they form.  There exist ideas as to why Mercury is so much smaller than Earth and Venus but they remain speculative and narrow.  Only recently was a cohesive theory presented to explain why Mars is smaller than Earth, and more work is needed to confirm or refute it.   

We first present the observational constraints in the Solar System and extra-solar planetary systems in \S 2.  Next, we review the dynamics of accretion of planetary embryos from planetesimals in \S 3, and of terrestrial planets from embryos in \S 4, including a discussion of the importance of a range of parameters.  In \S 5 we apply accretion models to extra-solar planets and in \S 6 to the Solar System.  We discuss different modes of accretion and current limitations in \S 7 and summarize in \S 8.

\section{\textbf{OBSERVATIONAL CONSTRAINTS}}

Given the explosion of new discoveries in extra-solar planets and our detailed knowledge of the Solar System, there are ample observations with which to constrain accretion models.  Given the relatively low resolution of numerical simulations, accretion models generally attempt to reproduce large-scale constraints such as planetary mass-and orbital distributions rather than smaller-scale ones like the exact characteristics of each planet.  We now summarize the key constraints for the Solar System and exoplanets.


\bigskip
\noindent
\textbf{2.1 The Solar System}
\bigskip

{\bf The masses and orbits of the terrestrial planets.}  There exist metrics to quantify different aspects of a planetary system and to compare it with the Solar System.  The angular momentum deficit $AMD$~\citep{laskar97} measures the difference in orbital angular momentum between the planets' orbits and the same planets on circular, coplanar orbits.  The $AMD$ is generally used in its normalized form:
\begin{equation} 
AMD = \frac{\sum_{j} m_j \sqrt{a_j} \left(1 - cos(i_j) \sqrt{1-e_j^2}\right)} {\sum_j m_j \sqrt{a_j}}, 
\end{equation}
\noindent where $a_j$, $e_j$, $i_j$, and $m_j$ are planet $j$'s semimajor axis, eccentricity, inclination and mass. The $AMD$ of the Solar System's terrestrial planets is 0.0018.  

The radial mass concentration $RMC$~\citep[defined as $S_c$ by][]{chambers01} measures the degree to which a system's mass is concentrated in a small region:
\begin{equation} 
RMC = max \left(\frac{\sum m_j}{\sum m_j [log_{10}(a/a_j)]^2} \right).
\end{equation}
Here, the function in brackets is calculated for $a$ across the planetary system, and the $RMC$ is the maximum of that function.  For a single-planet system the $RMC$ is infinite.  The $RMC$ is higher for systems in which the total mass is packed in smaller and smaller radial zones.  The $RMC$ is thus smaller for a system with equal-mass planets than a system in which a subset of planets dominate the mass.  The $RMC$ of the Solar System's terrestrial planets is 89.9.

{\bf The geochemically-determined accretion histories of Earth and Mars.}  Radiogenic elements with half-lives of a few to 100 Myr can offer concrete constraints on the accretion of the terrestrial planets.  Of particular interest is the $^{182}$Hf-$^{182}$W system, which has a half life of 9 Myr.  Hf is lithophile (``rock-loving") and W is siderophile (``iron-loving'').  The amount of W in a planet's mantle relative to Hf depends on the timing of core formation~\citep{nimmo06}.  Early core formation (also called ``core closure'') would strand still-active Hf and later its product W in the mantle, while late core formation would cause all W to be sequestered in the core and leave behind a W-poor mantle.  Studies of the Hf-W system have concluded that the last core formation event on Earth happened roughly 30-100 Myr after the start of planet formation~\citep{kleine02,yin02,kleine09,konig11}.  Similar studies on martian meteorites show that Mars' accretion finished far earlier, within 5 Myr~\citep{nimmo07,dauphas11}.  

The highly-siderophile element (HSE) contents of the terrestrial planets' mantles also provide constraints on the total amount of mass accreted by a planet after core closure~\citep{drake02}.  This phase of accretion is called the {\em late veneer}~\citep{kimura74}.  Several unsolved problems exist regarding the late veneer, notably the very high Earth/Moon HSE abundance ratio~\citep{day07,walker09}, which has been proposed to be the result of either a top-heavy~\citep{bottke10,raymond13} or bottom-heavy~\citep{schlichting12} distribution of planetesimal masses.

{\bf The large-scale structure of the asteroid belt.}  Reproducing the asteroid belt is not the main objective of formation models.  But any successful accretion model must be consistent with the asteroid belt's observed structure, and that structure can offer valuable information about planet formation.  Populations of small bodies can be thought of as the ``blood spatter on the wall'' that helps detectives solve the ``crime'', figuratively speaking of course.  

The asteroid belt's total mass is just $5 \times10^{-4} \mearth$, about four percent of a lunar mass.  This is 3-4 orders of magnitude smaller than the mass contained within the belt for any disk surface density profile with a smooth radial slope.  In addition, the inner belt is dominated by more volatile-poor bodies such as E-types and S-types whereas the outer belt contains more volatile-rich bodies such as C-types and D-types~\citep{gradie82,demeo13}.  There are no large gaps in the distribution of asteroids -- apart from the Kirkwood gaps associated with strong resonances with Jupiter -- and this indicates that no large ($\gtrsim 0.05 \mearth$) embryos were stranded in the belt after accretion, even if the embryos could have been removed during the late heavy bombardment~\citep{raymond09c}. 

{\bf The existence and abundance of volatile species -- especially water -- on Earth.}  Although it contains just 0.05-0.1\% water by mass~\citep{lecuyer98,marty12}, Earth is the wettest terrestrial planet.  It is as wet as ordinary chondrite meteorites, thought to represent the S-type asteroids that dominate the inner main belt, and wetter than enstatite chondrites that represent E-types interior to the main belt~\citep[see, for example, figure 5 from][]{morby12b}.  We think that this means that the rocky building blocks in the inner Solar System were dry.  In addition, heating mechanisms such as collisional heating and radiogenic heating from $^{26}$Al may have dehydrated fast-forming planetesimals~\citep[e.g.][]{grimm93}.  The source of Earth's water therefore requires an explanation. 

The isotopic composition of Earth's water constrains its origins.  The D/H ratio of water on Earth is a good match to carbonaceous chondrite meteorites thought to originate in the outer asteroid belt~\citep{marty06}.  The D/H of most observed comets is $2\times$ higher -- although one comet was recently measured to have the same D/H as Earth~\citep{hartogh11} -- and that of the Sun (and presumably the gaseous component of the protoplanetary disk) is $6\times$ smaller~\citep{geiss98}.  It is interesting to note that, while the D/H of Earth's water can be matched with a weighted mixture of material with Solar and cometary D/H, that same combination does not match the $^{15}$N/$^{14}$N isotopic ratio~\citep{marty06}.  Carbonaceous chondrites, on the other hand, match both measured ratios.  

\begin{table}
\center
\scriptsize
\title{\scriptsize Key inner Solar System Constraints}
\begin{tabular}{lc}
\hline
Angular momentum deficit $AMD$& 0.0018 \\
Radial Mass Concentration $RMC$& 89.9 \\
Mars' accretion timescale$^1$ & 3-5 Myr\\
Earth's accretion timescale$^2$ & $\sim 50$ Myr\\
Earth's late veneer$^3$ & $(2.5-10) \times 10^{-3} \mearth$\\
Total mass in asteroid belt & $5 \times 10^{-4} \mearth$\\
Earth's water content by mass$^4$ & $5\times 10^{-4} - 3\times 10^{-3}$ \\
\hline
\end{tabular}
\caption{\footnotesize $^1$\cite{dauphas11}. $^2$\cite{kleine09,konig11}.  $^3$\cite{day07,walker09}, see also \cite{bottke10,schlichting12,raymond13}.  $^4$\cite{lecuyer98,marty12}}
\normalsize
\end{table} 


The bulk compositions of the planets are another constraint.  For example, the core/mantle (iron/silicate) mass ratideo of the terrestrial planets ranges from 0.4 (Mars) to 2.1 (Mercury).  The bulk compositions of the terrestrial planets depend on several factors in addition to orbital dynamics and accretion: the initial compositional gradients of embryos and planetesimals, evolving condensation fronts, and the compositional evolution of bodies due to collisions and evaporation.  Current models for the bulk composition of terrestrial planets piggyback on dynamical simulations such as the ones discussed in sections 4-6 below~\citep[e.g.][]{bond10,carterbond12,elser12}.  These represent a promising avenue for future work.

\bigskip
\noindent
\textbf{2.2 Extrasolar Planetary Systems}
\bigskip

{\bf The abundance and large-scale characteristics of ``hot Super Earths"}.  These are the terrestrial exoplanets whose origin we want to understand.  Radial velocity and transit surveys have shown that roughly 30-50\% of main sequence stars host at least one planet with $M_p \lesssim 10 \mearth$ with orbital period $P \lesssim 85-100$~days~\citep{mayor11,howard10,howard12,fressin13}.  Hot super Earths are preferentially found in multiple systems~\citep[e.g.][]{udry07,lissauer11}.  These systems are in compact orbital configurations that are similar to the Solar System's terrestrial planets' as measured by the orbital period ratios of adjacent planets.  The orbital spacing of adjacent Kepler planet candidates is also consistent with that of the Solar System's planets when measured in mutual Hill radii~\citep{fang13}.  

\begin{figure}[h]
 \epsscale{0.99}
\plotone{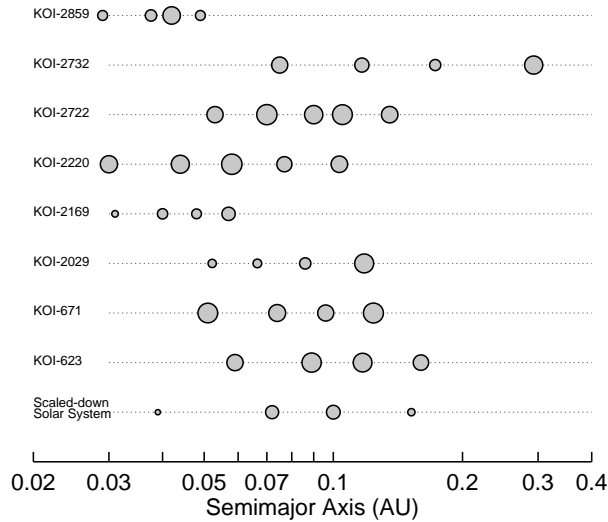}
\caption{\small Systems of (presumably) terrestrial planets.  The top 8 systems are candidate Kepler systems containing four or five planets that do not contain any planets larger than $1.5 {\rm R}_\oplus$~\citep[from][]{batalha13}.  The bottom system is the Solar System's terrestrial planets with semimajor axes scaled down by a factor of 10.  The size of each planet is scaled to its actual measured size (the Kepler planet candidates do not have measured masses). }  
\label{fig:kepler_ss}
\end{figure}

Figure~\ref{fig:kepler_ss} shows eight systems each containing 4-5 presumably terrestrial exoplanets discovered by the Kepler mission.  The largest planet in each system is less than 1.5 Earth radii, and in one system the largest planet is actually smaller than Earth (KOI-2169).  The Solar System is included for scale, with the orbit of each terrestrial planet shrunk by a factor of ten (but with their actual masses).  Given that the x axis is on a log scale, the spacing between planets is representative of the ratio between their orbital periods (for scale, the Earth/Venus period ratio is about 1.6).  

Given the uncertainties in the orbits of extra-solar planets and observational biases that hamper the detection of low-mass, long-period planets we do not generally apply the $AMD$ and $RMC$ metrics to these systems.  Rather, the main constraints come from the systems' orbital spacing, masses and mass ratios.  

{\bf The existence of giant planets on exotic orbits.}  Simulations have shown in planetary systems with giant planets the giants play a key role in shaping the accretion of terrestrial planets~\citep[e.g.,][]{chambers02,levison03,raymond04}.  Giant exoplanets have been discovered on diverse orbits that indicate rich dynamical histories.  Gas giants exist on orbits with eccentricities as high as 0.9.  It is thought that these planets formed in systems with multiple gas giants that underwent strong dynamical instabilities that ejected one or more planets and left behind surviving planets on eccentric orbits~\citep{chatterjee08,juric08,raymond10}.  Hot Jupiters -- gas giants very close to their host stars -- are thought to have either undergone extensive inward gas-driven migration~\citep{lin96} or been re-circularized by star-planet tidal interactions from very eccentric orbits produced by planet-planet scattering~\citep{nagasawa08,beauge12} or other mechanisms~\citep[e.g.][see chapter by Davies et al]{fabrycky07,naoz11}.  There also exist gas giants on nearly-circular Jupiter-like orbits~\citep[e.g.][]{wright08}.  However, from the current discoveries systems of gas giants like the Solar System's -- with giant planets confined beyond 5 AU on low-eccentricity orbits -- appear to be the exception rather than the rule.

Of course, many planetary systems do not host currently-detected giant planets.  Radial velocity surveys show that at least 14\% of Sun-like stars have gas giants with orbits shorter than 1000 days~\citep{mayor09}, and projections to somewhat larger radii predict that $\sim20\%$ have gas giants within 10 AU~\citep{cumming08}.  Although they are limited by small number statistics, the statistics of high-magnification (planetary) microlensing events suggest that 50\% or more of stars have gas giants on wide orbits~\citep{gould10}.  In addition, the statistics of short-duration microlensing events suggests that there exists a very abundant population of gas giants on orbits that are separated from their stars; these could either be gas giants on orbits larger than $\sim 10$~AU or free-floating planets~\citep{sumi11}.  

{\bf The planet-metallicity correlation.}  Gas giants -- at least those easily detectable with current techniques -- are observed to be far more abundant around stars with high metallicities~\citep{gonzalez97,santos01,laws03,fischer05}.  However, this correlation does not hold for low-mass planets, which appear to be able to form around stars with a wide range of metallicities~\citep{ghezzi10,buchhave12,mann12}.  It is interesting to note that there is no observed trend between stellar metallicity and the presence of debris disks~\citep{greaves06,moromartin07}, although disks do appear to dissipate faster in low-metallicity environments~\citep{yasui09}.  The planet-metallicity correlation in itself does strongly constraint the planet formation models we discuss here.  What is important is that the formation of systems of hot Super Earths does not appear to depend on the stellar metallicity, i.e. the solids-to-gas ratio in the disk.  



Additional constraints on the initial conditions of planet formation come from observations of protoplanetary disks around other stars~\citep{williams11}.  These observations measure the approximate masses and radial surface densities of planet-forming disks, mainly in their outer parts.  They show that protoplanetary disks tend to have masses on the order of $10^{-3}$-$10^{-1}$ times the stellar mass~\citep[e.g.][]{scholz06,andrews07a,eisner08,eisner12}, with typical radial surface density slopes of $\Sigma \propto r^{-(0.5-1)}$ in their outer parts~\citep{mundy00,looney03,andrews07b}.  In addition, statistics of the disk fraction in clusters with different ages show that the gaseous component of disks dissipate within a few Myr~\citep{haisch01,hillenbrand08,fedele10}.  It is also interesting to note that disks appear to dissipate more slowly around low-mass stars than Solar-mass stars~\citep{pascucci09}.  

\section{\textbf{FROM PLANETESIMALS TO PLANETARY EMBRYOS}}

In this section we summarize the dynamics of accretion of planetary embryos.  We first present the standard model of runaway and oligarchic growth from planetesimals (\S 3.1).  We next present a newer model based on the accretion of small pebbles (\S 3.2).

\bigskip
\noindent
\textbf{3.1 Runaway and Oligarchic Growth}
\bigskip

\smallskip
\noindent{\bf Growth Modes}

There are two growth modes: ``orderly'' and ``runaway''. 
In orderly growth, all planetesimals grow at the same rate, so the mass ratios between planetesimals tend to unity.
During runaway growth, on the other hand, larger planetesimals grow faster than smaller ones and mass ratios increase monotonically.  Consider the evolution of the mass ratio between two planetesimals with masses $M_1$ and $M_2$, assuming $M_1>M_2$.  The time derivative of the mass ratio is given by
\begin{equation}
\frac{\mathrm{d}}{\mathrm{d}t}\left(\frac{M_1}{M_2}\right) =
\frac{M_1}{M_2}
\left(
\frac{1}{M_1}\frac{\mathrm{d}M_1}{\mathrm{d}t} -
\frac{1}{M_2}\frac{\mathrm{d}M_2}{\mathrm{d}t} 
\right).
\end{equation}
It is the relative growth rate $(1/M)\mathrm{d}M/\mathrm{d}t$ that determines the growth mode.  If the relative growth rate decreases with $M$, $\mathrm{d}(M_1/M_2)/\mathrm{d}t$ is negative then the mass ratio tends
 to be unity.  This corresponds to orderly growth.  If the relative growth rate increases with $M$,
 $\mathrm{d}(M_1/M_2)/\mathrm{d}t$ is positive and the mass ratio increases, leading to runaway growth. 

The growth rate of a planetesimal with mass $M$ and radius $R$ that is accreting field planetesimals with mass $m$ ($M>m$) can be written as
\begin{equation}
\frac{\mathrm{d}M}{\mathrm{d}t} \simeq
n_m\pi R^2
\left(1+\frac{v_{\rm esc}^2}{v_{\rm rel}^2}\right)
v_{\rm rel}m,
\label{eq:growth_rate}
\end{equation}
where $n_m$ is the number density of field planetesimals, and $v_{\rm rel}$ and $v_{\rm esc}$ are the relative velocity between the test and the field planetesimals and the escape velocity from the surface of the test planetesimal, respectively \citep[e.g.,][]{kokubo96}.  The term $v_{\rm esc}^2/v_{\rm rel}^2$ indicates the enhancement of collisional cross-section by gravitational focusing.

\smallskip
\noindent{\bf Runaway Growth of Planetesimals}



The first dramatic stage of accretion through which a population of planetesimals passes is runaway growth \citep{greenberg78,wetherill89,kokubo96}.  During planetesimal accretion gravitational focusing is efficient because the velocity dispersion of planetesimals is kept smaller than the escape velocity due to gas drag. In this case Eq.\ref{eq:growth_rate} reduces to
\begin{equation}
\label{eq:rdmdt}
\frac{\mathrm{d}M}{\mathrm{d}t} \propto \Sigma_\mathrm{dust} M^{4/3}v^{-2},
\end{equation}
 where $\Sigma_\mathrm{dust}$ and $v$ are the surface density and velocity dispersion of planetesimals and we used $n_m\propto \Sigma_\mathrm{dust} v^{-1}$, $v_{\rm esc} \propto M^{1/3}$, $R \propto M^{1/3}$, and $v_{\rm rel}\simeq v$.  During the early stages of accretion, $\Sigma_\mathrm{dust}$ and $v$ barely depend on $M$, in other words, the reaction of growth on $\Sigma_\mathrm{dust}$ and $v$ can be neglected since the mass in small planetesimals dominate the system.   
In this case we have 
\begin{equation}
\frac{1}{M}\frac{\mathrm{d}M}{\mathrm{d}t} \propto M^{1/3},
\end{equation}
which leads to runaway growth.

During runaway growth, the eccentricities and inclinations of the largest bodies are kept small by dynamical friction from smaller bodies \citep{wetherill89,ida92}.  Dynamical friction is an equipartitioning of energy that maintains lower random velocities -- and therefore lower-eccentricity and lower-inclination orbits -- for the largest bodies.  The mass distribution relaxes to a distribution that is well approximated by a power-law distribution.  Among the large bodies that form in simulations of runaway growth, the mass follows a distribution $\mathrm{d}n_{\rm c}/\mathrm{d}m \propto m^{y}$, where $y \simeq -2.5$.  This index can be derived analytically as a stationary distribution~\citep{makino98}.   The power index smaller than -2 is characteristic of runaway growth, as most of the system mass is contained in small bodies.  We also note that runaway growth does not necessarily mean that the growth time decreases with mass, but rather that the mass ratio of any two bodies increases with time.

\smallskip
\noindent{\bf Oligarchic Growth of Planetary Embryos}

During the late stages of runaway growth, embryos grow while interacting with one another.  The dynamics of the system become dominated by a relatively small number -- a few tens to a few hundred -- oligarchs~\citep{kokubo98,kokubo00,thommes03}.  

Oligarchic growth is the result of the self-limiting nature of runaway growth and orbital repulsion of planetary embryos. The formation of similar-sized planetary embryos is due to a slow-down of runaway growth \citep{lissauer87,ida93,ormel10b}.  When the mass of a planetary embryo $M$ exceeds about 100 times that of the average planetesimal, the embryo increases the random velocity of neighboring planetesimals to be $v \propto M^{1/3}$~\citep[but note that this depends on the planetesimal size;][]{ida93,rafikov04,chambers06}.  The relative growth rate (from Eq.\ref{eq:rdmdt}) becomes
\begin{equation}
\frac{1}{M}\frac{\mathrm{d}M}{\mathrm{d}t} \propto \Sigma_\mathrm{dust} M^{-1/3}.
\end{equation}
$\Sigma_\mathrm{dust}$ decreases through accretion of planetesimals by the embryo as $M$ increases \citep{lissauer87}.  The relative growth rate is a decreasing function of $M$, which changes the growth mode to orderly.  Neighboring embryos grow while maintaining similar masses.  During this stage, the mass ratio of an embryo to its neighboring planetesimals increases because for the planetesimals with mass $m$, 
 $(1/m)\mathrm{d}m/\mathrm{d}t \propto \Sigma_\mathrm{dust} m^{1/3}M^{-2/3}$, such that
\begin{equation}
\frac{(1/M)\mathrm{d}M/\mathrm{d}t}{(1/m)\mathrm{d}m/\mathrm{d}t} \propto \left(\frac{M}{m}\right)^{1/3}.
\end{equation}
The relative growth rate of the embryo is by a factor of $(M/m)^{1/3}$ larger than the planetesimals'.  A bi-modal  embryo-planetesimal system is formed.  While the planetary embryos grow, a process called orbital repulsion keeps their orbital separations at roughly 10 mutual Hill radii $R_{H,m}$, where $R_{H,m} = 1/2 \left(a_1+a_2\right) \left[(M_1+M_2)/ (3 M_\star)\right]^{1/3}$; here subscripts 1 and 2 refer to adjacent embryos. Orbital repulsion is a coupling effect of gravitational scattering between planetary embryos that increases their orbital separation and eccentricities and dynamical friction from small planetesimals that decreases the eccentricities \citep{kokubo95}.  Essentially, if two embryos come too close to each other their eccentricities are increased by gravitational perturbations.  Dynamical friction from the planetesimals re-circularizes their orbits at a wider separation.  

\begin{figure}[h]
 \includegraphics[width=0.45\textwidth]{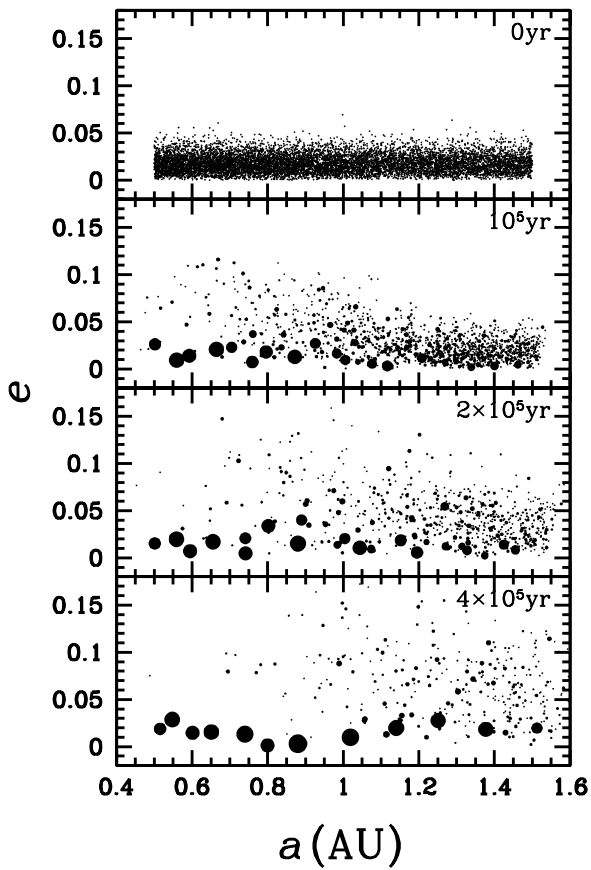}
 \caption{Oligarchic growth of planetary embryos.  Snapshots of the planetesimal system on the $a$-$e$ plane are shown for $t=0$, $10^5$, $2\times 10^5$, and $4\times 10^5$ years. The circles represent planetesimals with radii proportional to their true values. The initial planetesimal system consists of 10000 equal-mass ($m = 2.5 \times 10^{-4} \mearth$) bodies. In this simulation, a 6-fold increase in the planetesimal radius was used to accelerate accretion.  In $4\times 10^5$ years, the number of bodies decreases to 333.  From~\cite{kokubo02}. 
 }
 \label{fig:a-e-t_oligarchic}
\end{figure}

An example of oligarchic growth is shown in Figure~\ref{fig:a-e-t_oligarchic} \citep{kokubo02}.  About 10 embryos form with masses comparable to Mars' ($M \approx 0.1 \mearth$) on nearly circular non-inclined orbits with characteristic orbital separations of $10 R_{H,m}$ .  At large $a$ the planetary embryos are still growing at the end of the simulation.

Although oligarchic growth describes the accretion of embryos from planetesimals, it implies giant collisions between embryos that happen relatively early and are followed by a phase of planetesimal accretion.  Consider the last pairwise accretion of a system of oligarchs on their way to becoming planetary embryos.  The oligarchs have masses $M_{olig}$ and are spaced by $N$ mutual Hill radii $R_{H,m}$, where $N \approx 10$ is the rough stability limit for such a system.  The final system of embryos will likewise be separated by $N \, R_{H,m}$, but with larger masses $M_{emb}$.  The embryos grow by accreting material within an annulus defined by the inter-embryo separation.  Assuming pairwise collisions between equal-mass oligarchs to form a system of equal-mass embryos, the following simple relation should hold: $N R_{H,m} (M) = 2 \,N \,R_{H,m}(M_{emb})$.  Given that $R_{H,m}(M) \sim (2M)^{1/3}$, this implies that $M_{emb} = 8 M_{olig}$.  After the collision between a pair of oligarchs, each embryo must therefore accrete the remaining three quarters of its mass from planetesimals.  

We can estimate the dynamical properties of a system of embryos formed by oligarchic growth.  We introduce a protoplanetary disk with surface density of dust and gas $\Sigma_\mathrm{dust}$ and $\Sigma_\mathrm{gas}$ defined as:
\begin{eqnarray}
\Sigma_\mathrm{dust} = f_\mathrm{ice}\Sigma_1\scale{a}{1\,{\rm AU}}{-x} \ {\rm gcm}^{-2}  \nonumber \\
\Sigma_\mathrm{gas}  = f_\mathrm{gas}\Sigma_1\scale{a}{1\,{\rm AU}}{-x} \ {\rm gcm}^{-2},
\end{eqnarray}
where $\Sigma_1$ is simply a reference surface density in solids at 1 AU and $x$ is the radial exponent.  $f_\mathrm{ice}$ and $f_\mathrm{gas}$ are factors that enhance the surface density of ice and gas with respect to dust.  In practice $f_\mathrm{ice}$ is generally taken to be 2-4~\citep[see][]{kokubo02,lodders03} and $f_\mathrm{gas}\approx 100$.  Given an orbital separation $b$ of embryos, the isolation (final) mass of a planetary embryo at orbital radius $a$ is estimated as \citep{kokubo02}:
\begin{eqnarray}
\label{eq:m_iso}
M_\mathrm{iso} & \simeq 2\pi a b \Sigma_\mathrm{dust} = 0.16
\scale{b}{10 r_\mathrm{H}}{3/2}
\scale{f_\mathrm{ice}\Sigma_1}{10}{3/2}\nonumber \\
&
\scale{a}{1\,{\rm AU}}{(3/2)(2-x)}
\scale{M_\star}{M_\odot}{-1/2}
M_\oplus, 
\end{eqnarray}
where $M_{\star}$ is the stellar mass. The time evolution of an oligarchic body is~\citep{thommes03,chambers06}:
\begin{equation}
M(t)= M_{\rm iso}\tanh^3\left(\frac{t}{\tau_{\rm grow}} \right). \label{eq:me2}
\end{equation}
The growth timescale $\tau_{\rm grow}$ is estimated as 
\begin{eqnarray}
\tau_{\rm grow} &=& 1.1 \times 10^{6} f_{\rm ice}^{-1/2} 
\left(\frac{f_{\rm gas}}{240}\right)^{-2/5}
\left(\frac{\Sigma_1}{10}\right)^{-9/10} \nonumber \\
& & \left(\frac{b}{10 r_{\rm H}}\right)^{1/10}
\left(\frac{a}{\rm 1 \hspace{0.3em} AU}\right)^{8/5+9x/10}
 \left(\frac{M_{\star}}{M_{\odot}}\right)^{-8/15} \nonumber \\
& & \left(\frac{\rho_{\rm p}}{2 \hspace{0.3em} {\rm gcm}^{-3}}\right)^{11/15} 
\left(\frac{r_{\rm p}}{100 \hspace{0.3em} {\rm km}}\right)^{2/5} \hspace{0.3em} {\rm yr},
\end{eqnarray}
where $r_{\rm p}$ and $\rho_{\rm p}$ are the physical radius and internal density of planetesimals.
Eq.~(\ref{eq:me2}) indicates that the embryo gains 44$\%$, 90$\%$, and  99$\%$ of its final mass during 
1$\tau_{\rm grow}$, 2$\tau_{\rm grow}$, and 3$\tau_{\rm grow}$.  
  
For the standard disk model defined above, $M_\mathrm{iso} \sim 0.1 \mearth$ in the terrestrial planet region. This suggests that if they formed by oligarchic growth, Mercury and Mars may simply represent leftover planetary embryos.  A short growth timescale ($\tau_{\rm grow} < 2$ Myr) of Mars estimated by  the Hf-W chronology~\citep{dauphas11} would suggest that Mars accreted from a massive disk of small planetesimals~\citep{kobayashi13,morishima13}.  Alternately, accretion of larger planetesimals might have been truncated as proposed by the Grand Tack model (see \S 6.3).  Unlike Mars and Mercury, further accretion of planetary embryos is necessary to complete Venus and Earth. This next, final stage is called late-stage accretion (see Section 4).



\bigskip
\noindent
\textbf{3.2 Embryo formation by pebble accretion}
\bigskip


\cite{lambrechts12}, hereafter LJ12, proposed a new model of growth for planetary embryos and giant planet cores.  They argued that if the disk's mass is dominated by pebbles of a few decimeters in size, the largest planetesimals accrete pebbles very efficiently and can rapidly grow to several Earth masses~\cite[see also][]{johansen10,ormel10,murrayclay11}.  This model builds on a recent planetesimal formation model in which large planetesimals (with sizes from $\sim 100$ up to $\sim$1,000km) form by the collapse of a self-gravitating clump of pebbles, concentrated to high densities by disk turbulence and the streaming instability~\citep[][see also chapter by Johansen et al]{youdin05,johansen06,johansen07,johansen09}.  The pebble accretion scenario essentially describes how large planetesimals continue to accrete.  There is observational evidence for the existence of pebble-sized objects in protoplanetary disks~\citep{wilner05,rodmann06,lommen07,perez12}, although their abundance relative to larger objects (planetesimals) is unconstrained.

Pebbles are strongly coupled with the gas so they encounter the already-formed planetesimals with a velocity $\Delta v$ that is equal to the difference between the Keplerian velocity and the orbital velocity of the gas, which is slightly sub-Keplerian due to the outward pressure gradient. LJ12 define the planetesimal {\it Bondi radius} as the distance at which the planetesimal exerts a deflection of one radian on a particle approaching with a velocity $\Delta v$:
\begin{equation}
R_B={{GM}\over{\Delta v^2}}
\label{Bondi}
\end{equation}
where $G$ is the gravitational constant and $M$ is the planetesimal mass (the deflection is larger if the particle passes closer than $R_B$). LJ12 showed that all pebbles with a stopping time $t_f$ smaller than the Bondi time $t_B=R_B/\Delta v$ that pass within a distance $R=(t_f/t_B)^{1/2} R_B$ spiral down towards the planetesimal and are accreted by it. Thus, the growth rate of the planetesimal is:
\begin{equation}
\mathrm{d}M/\mathrm{d}t=\pi\rho R^2\Delta v
\label{Bondi-accrete}
\end{equation}
where $\rho$ is the volume density of the pebbles in the disk. Because $R\propto M$, the accretion rate $\mathrm{d}M/\mathrm{d}t\propto M^2$. Thus, pebble accretion is at the start a super-runaway process that is faster than the runaway accretion scenario (see Sec 3.1) in which $\mathrm{d}M/\mathrm{d}t\propto M^{4/3}$. According to LJ12, this implies that in practice, only planetesimals more massive than $\sim 10^{-4}\mearth$ (comparable to Ceres' mass) undergo significant pebble accretion and can become embryos/cores.

The super-runaway phase cannot last very long. When the Bondi radius exceeds the scale height of the pebble layer, the accretion rate becomes
\begin{equation}
\mathrm{d}M/\mathrm{d}t=2R\Sigma \Delta v
\end{equation}
where $\Sigma$ is the surface density of the pebbles. This rate is proportional to $M$, at the boundary between runaway and orderly (oligarchic) growth.  

\begin{figure}[h]
 \includegraphics[width=0.45\textwidth]{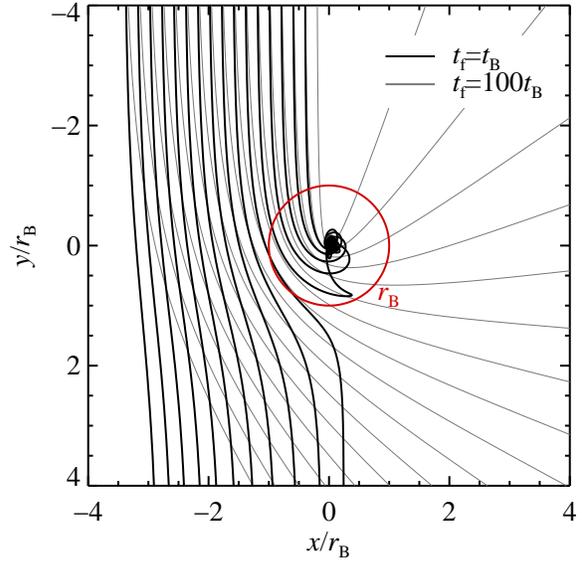}
 \caption{Trajectories of particles in the vicinity of a growing embryo.  The black curves represent particles strongly coupled to the gas and the gray curves particles that are weakly coupled, as measured by the ratio of the stopping time $t_f$ to the Bondi time $t_B$.  The orbits of weakly-coupled particles are deflected by the embryo's gravity, but the strongly coupled particles spiral inward and are quickly accreted onto the embryo.  From~\cite{lambrechts12}.}
 \label{fig:pebble}
\end{figure}

Moreover, when the Bondi radius exceeds the Hill radius $R_H = a \left[M/(3 M_\star)\right]^{1/3}$, the accretion rate becomes
\begin{equation}
\mathrm{d}M/\mathrm{d}t=2R_H \Sigma v_H
\label{Hill-accretion}
\end{equation}
where $v_H$ is the Hill velocity (i.e. the difference in Keplerian velocities between two circular orbits separated by $R_H$).  Here $\mathrm{d}M/\mathrm{d}t \propto M^{2/3}$ and pebble accretion enters an oligarchic regime.

For a given surface density of solids $\Sigma$ the growth of an embryo is much faster if the solids are pebble-sized than planetesimal sized. This is the main advantage of the pebble-accretion model. However, pebble accretion ends when the gas disappears from the protoplanetary disk, whereas runaway/oligarchic accretion of planetesimals can continue.  Also, the ratio between $\Sigma_{\rm planetesimals}/\Sigma_{\rm pebbles}$ remains to be quantified, and ultimately it is this ratio that determines which accretion mechanism is dominant.

An important problem in Solar System formation is that the planetary embryos in the inner solar system are thought to have grown only up to at most a Mars-mass, whereas in the outer solar system some of them reached many Earth masses, enough to capture a primitive atmosphere and become giant planets. The difference between these masses can probably be better understood in the framework of the pebble-accretion model than in the planetesimal-accretion model. 

The dichotomy in embryo mass in the inner/outer Solar System might have been caused by radial drift of pebbles.  We consider a disk with a ``pressure bump''~\citep{johansen09} at a given radius $R_{\rm bump}$.  At this location the gas' azimuthal velocity $v_\theta$ is larger than the Kepler velocity $v_K$.  Pebbles cannot drift from beyond $R_{\rm bump}$to within $R_{\rm bump}$ because they are too strongly coupled to the gas. Embryos growing interior to $R_{\rm bump}$ are thus ``starved'' in the sense that they can only accrete pebbles within $R_{\rm bump}$, and are not in contact with the presumably much larger pebble reservoir beyond $R_{\rm bump}$. Of course, embryos growing exterior to $R_{\rm bump}$ would not be starved and could grow much faster and achieve much larger masses within the gaseous disk's lifetime. On the contrary, the planetesimal accretion model does not seem to present a sharp radial boundary for slow/fast accretion and so it is harder to understand the dichotomy of embryo masses in that framework.  

\cite{ida08} argued that a pressure bump could be located at the snow line. If this is true, then we can speculate that giant planet cores should form in the icy part of the disk and sub-Mars-mass planetary embryos in the rocky part of the disk.  This seems to be consistent with the structure of the Solar System.

\section{\textbf{FROM PLANETARY EMBRYOS TO TERRESTRIAL PLANETS}}

The final accumulation of terrestrial planets -- sometimes called late-stage accretion -- is a chaotic phase characterized by giant embryo-embryo collisions.  It is during this phase that the main characteristics of the planetary system are set: the planets' masses and orbital architecture, the planets' feeding zones and thus their bulk compositions, and their spin rates and obliquities~\citep[although their spins may be altered by other processes on long timescales -- see e.g.,][]{correia09}.  

Whether embryos form by accreting planetesimals or pebbles, the late evolution of a system of embryos is likely in the oligarchic regime.  The transition from oligarchic growth to late-stage accretion happens when there is insufficient damping of random velocities by gas drag and dynamical friction from planetesimals~\citep{kenyon06}.  The timescale of the orbital instability of an embryo system has been numerically calculated by $N$-body simulations to be
\begin{equation}
\log t_\mathrm{inst} \simeq
 c_1 \left(\frac{b_\mathrm{ini}}{r_\mathrm{H}}\right) + c_2,
\end{equation}
where $b_\mathrm{ini}$ is the initial orbital separation of adjacent embryos and $c_1$ and $c_2$ are functions of the initial $\langle e^2\rangle^{1/2}$ and $\langle i^2\rangle^{1/2}$ of the system~\citep{chambers96,yoshinaga99}.  

The most important quantity in determining the outcome of accretion is the level of eccentricity excitation of the embryos.  This is determined by a number of parameters including forcing from any giant planets that exist in the system~\citep{chambers02,levison03,raymond04}.  Although giant planets are far larger than terrestrials, they are thought to form far faster and strongly influence late-stage terrestrial accretion.  The lifetimes of gaseous protoplanetary disks are just a few Myr~\citep{haisch01} whereas geochemical constraints indicate that Earth took 50-100 Myr to complete its formation~\citep{touboul07,kleine09,konig11}.  The dynamics described in this section are assumed to occur in a gas-free environment (we consider the effects of gas in other sections). 

We first describe the dynamics of accretion and radial mixing (\S 4.1), then the effect of accretion on the final planets' spins (\S4.2) and the effect of embryo and disk parameters on accretion (\S4.3).  We explain the consequences of taking into account imperfect accretion (\S4.4) and the effect of giant planets on terrestrial accretion (\S4.5).  

\bigskip
\noindent
\textbf{4.1 Timescales and Radial Mixing}
\bigskip

\begin{figure}[h]
 \epsscale{1}
\plotone{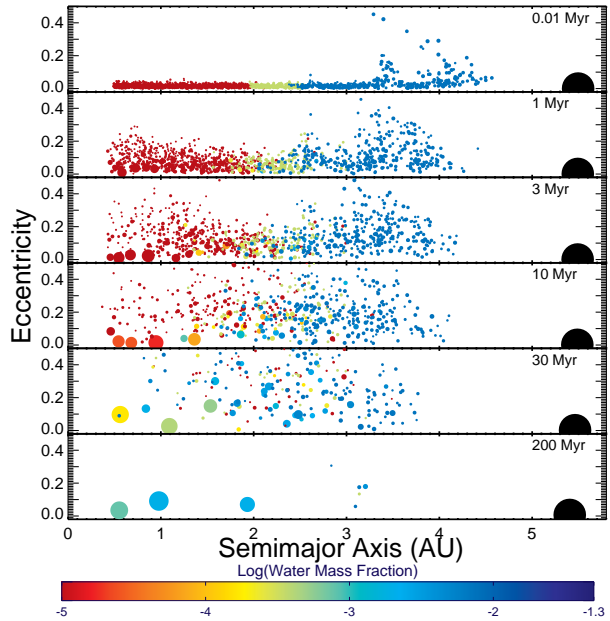}
\caption{\small Six snapshots of a simulation of terrestrial planet formation~\citep[adapted from][]{raymond06b}.  The simulation started from 1885 self-gravitating sub-lunar-mass bodies spread from 0.5 to 5 AU following an $r^{-3/2}$ surface density profile, comprising a total of $9.9\mearth$.  The large black circle represents a Jupiter-mass planet.  The size of each body is proportional to its mass$^{1/3}$.  The color represents each body's water content (see color bar).}  
\label{fig:sim0}
\end{figure}

Figure~\ref{fig:sim0} shows the evolution of a simulation of late-stage accretion from~\cite{raymond06b} that included a single Jupiter-mass giant planet on a circular orbit at 5.5 AU.  The population of embryos is excited from the inside-out by mutual scattering among bodies and from the outside-in by secular and resonant excitation by the giant planet.  Accretion is faster closer-in and proceeds as a wave sweeping outward in time.  At 10 Myr the disk inside 1 AU is dominated by 4 large embryos with masses close to Earth's.  The population of close-in (red) planetesimals has been strongly depleted, mainly by accretion but also by some scattering to larger orbital radii.  Over the rest of the simulation the wave of accretion sweeps outward across the entire system.  Small bodies are scattered onto highly-eccentric orbits and either collide with growing embryos or venture too close to the giant planet and are ejected from the system.  Embryos maintain modest eccentricities by dynamical friction from the planetesimals.  Nonetheless, strong embryo-embryo gravitational scattering events spread out the planets and lead to giant impacts such as the one thought to be responsible for creating Earth's Moon~\citep{cuk12,canup12}.  

After 200 Myr three terrestrial planets remain in the system with masses of 1.54, 2.04, and $0.95 \mearth$ (inner to outer).  Although modestly more massive, the orbits of the two inner planets are decent analogs for Earth and Venus.  The outer planet does a poor job of reproducing Mars: it is nine times too massive and too far from the star.  This underscores the {\em small Mars} problem: simulations that do not invoke strong external excitation of the embryo swarm systematically produce Mars analogs that are far too massive~\citep{wetherill91,raymond09c}.  We will return to this problem in \S 6.

A large reservoir of water-rich material is delivered to the terrestrial planets in the simulation from Fig.~\ref{fig:sim0}.  By 10 Myr four large embryos have formed inside 1 AU but they remain dry because to this point their feeding zones have been restricted to the inner planetary system.  Over the following 20 Myr planetesimals and embryos from the outer planetary system are scattered inward by repeated gravitational encounters with growing embryos.  These bodies sometimes collide with the growing terrestrial planets.  This effectively widens the feeding zones of the terrestrial planets to include objects that condensed at different temperatures and therefore have different initial compositions~\citep[see also][]{bond10,carterbond12,elser12}.  The compositions of the terrestrial planets become mixtures of the compositions of their constituent embryos and planetesimals.  The planets' feeding zones represent those constituents.  When objects from past 2.5 AU are accreted, water-rich material is delivered to the planet in the form of hydrated embryos and planetesimals.  In the simulations, from 30-200 Myr the terrestrial planets accrete objects from a wide range of initial locations and are delivered more water.  

Given that the water delivered to the planets in this simulation originated in the region between 2.5 and 4 AU, its composition should be represented by carbonaceous chondrites, which provide a very good match to Earth's water~\citep{morby00,marty06}.  The planets are delivered a volume of water that may be too large.  For example, the Earth analog's final water content by mass was $8 \times 10^{-3}$, roughly 8-20 times the actual value.  However, water loss during giant impacts was not taken into account in the simulation~\citep[see, e.g., ][]{genda05}.

\bigskip
\noindent
\textbf{4.2 Planetary spins}
\bigskip

Giant impacts impart large amounts of spin angular momentum on the terrestrial planets \citep[e.g.,][]{safronov69,lissauer91,dones93}.  The last few giant impacts tend to dominate the spin angular momentum~\citep{agnor99, kokubo07, kokubo10}.  Using a ``realistic'' accretion condition of planetary embryos~\citep[][see \S 4.4)]{genda12}, \cite{kokubo10} found that the spin angular velocity of accreted terrestrial planets follows a Gaussian distribution with a nearly mass-independent average value of about 70\% of the critical angular velocity for rotational breakup 
\begin{equation}
\omega_{\rm cr} = \left(\frac{GM}{R^3}\right)^{1/2}, 
\end{equation} 
where $M$ and $R$ are the mass and radius of a planet.  This appears to be a natural outcome of embryo-embryo impacts at speeds slightly larger than escape velocity.  At later times, during the late veneer phase, the terrestrial planets' spins are further affected by impacts with planetesimals~\citep{raymond13}.  



The obliquity of accreted planets ranges from 0$^\circ$ to 180$^\circ$ and follows an isotropic distribution \citep{agnor99, kokubo07, kokubo10}.  Both prograde and retrograde spins are equally probable. The isotropic distribution of $\varepsilon$ is a natural outcome of giant impacts. During the giant impact stage, the thickness of a planetary embryo system is $\sim a\langle i^2\rangle^{1/2} \sim 10r_{\rm H}$, far larger than the radius $R$ of planetary embryos $R\sim 10^{-2}r_{\rm H}$, where $a$, $i$, and $r_\mathrm{H}$ are the semimajor axis, inclination and Hill radius of planetary embryos.  Thus, collisions are fully three-dimensional and isotropic, which leads to isotropic spin angular momentum. This result clearly shows that prograde spin with small obliquity, which is common to the terrestrial planets in the solar system except for Venus, is not a common feature for planets assembled by giant impacts. Note that the initial obliquity of a planet determined by giant impacts can be modified substantially by stellar tide if the planet is close to the star and by satellite tide if the planet has a large satellite. 

\bigskip
\noindent
\textbf{4.3 Effect of disk and embryo parameters}
\bigskip

%

The properties of a system of terrestrial planets are shaped in large part by the total mass and mass distribution within the disk, and the physical and orbital properties of planetary embryos and planetesimals within the disk.  However, while certain parameters have a strong impact on the outcome, others have little to no effect.  

\cite{kokubo06} performed a suite of simulations of accretion of populations of planetary embryos to test the importance of the embryo density, mass, spacing and number.  They found that the bulk density of the embryos had little to no effect on the accretion within the range that they tested, $\rho = 3.0-5.5 {\rm g \, cm^{-3}}$.  One can imagine that the dynamics could be affected for extremely high values of $\rho$, if the escape speed from embryos were to approach a significant fraction of the escape speed from the planetary system~\citep{goldreich04}.  In practice this is unlikely to occur in the terrestrial planet forming region because it would require unphysically-large densities.  The initial spacing likewise had no meaningful impact on the outcome, at least when planetary embryos were spaced by 6-12 mutual Hill radii~\citep{kokubo06}.  Likewise, for a fixed total mass in embryos, the embryo mass was not important.

The total mass in embryos does affect the outcome.  A more massive disk of embryos and planetesimals produces fewer, more massive planets than a less massive disk~\citep{kokubo06,raymond07b}.  Embryos' eccentricities are excited more strongly in massive disks by encounters with massive embryos.  With larger mean eccentricities, the planets' feeding zones are wider than if the embryos' eccentricities were small, simply because any given embryos crosses a wider range of orbital radii.  The scaling between the mean accreted planet mass and the disk mass is therefore slightly steeper than linear: the mean planet mass $M_p$ scales with the local surface density $\Sigma_0$ as $M_p \propto \Sigma_0^{1.1}$~\citep{kokubo06}.  It is interesting to note that this scaling is somewhat shallower than the $\Sigma_0^{1.5}$ scaling of embryo mass with the disk mass~\citep{kokubo00}.  Accretion also proceeds faster in high-mass disks, as the timescale for interaction drops.  

Terrestrial planets that grow from a disk of planetesimals and planetary embryos retain a memory of the surface density profile of their parent disk.  In addition, the dynamics is influenced by which part of the disk contains the most mass.  In disks with steep density profiles -- i.e., if the surface density scales with orbital radius as $\Sigma \propto r^{-x}$, disks with large values of $x$ -- more mass is concentrated in the inner parts of the disk, where the accretion times are faster and protoplanets are dry.  Compared with disks with shallower density profiles (with small $x$), in disks with steep profiles the terrestrial planets tend to be more massive, form more quickly, form closer-in, and contain less water~\citep{raymond05b,kokubo06}.  

\bigskip
\noindent
\textbf{4.4 Effect of imperfect accretion}
\bigskip

As planetesimalsÕ eccentricities are excited by growing embryos, they undergo considerable collisional grinding. Collisional disruption can be divided into two types: catastrophic disruption due to high-energy impacts and cratering due to low-energy impacts.  \cite{kobayashi10b} found that cratering collisions are much more effective in collisional grinding than collisions causing catastrophic disruption, simply because the former impacts occur much more frequently than the latter ones. Small fragments are easily accreted by embryos in the presence of nebular gas~\citep{wetherill93}, although they rapidly drift inward due to strong gas drag, leading to small embryo masses~\citep{chambers08,kobayashi10}. 


Giant impacts between planetary embryos often do not result in net accretion.  Rather, there exists a diversity of collisional outcomes.  These include near-perfect merging at low impact speeds and near head-on configurations, partial accretion at somewhat higher impact speeds and angles, ``hit and run'' collisions at near-grazing angles, and even net erosion for high-speed, near head-on collisions~\citep{agnor04,asphaug06,asphaug10}.  Two recent studies used large suites of SPH simulations to map out the conditions required for accretion in the parameter space of large impacts~\citep{genda12,leinhardt12}.  However, most $N$-body simulations of terrestrial planet formation to date have assumed perfect accretion in which all collisions lead to accretion.


%
 
About half of the embryo-embryo impacts in a typical simulation of late-stage accretion do not lead to net growth~\citep{agnor04,kokubo10}.  Rather, the outcomes are dominated by partially accreting collision, hit-and-run impacts, and graze-and-merge events in which two embryos dissipate sufficient energy during a grazing impact to become gravitationally bound and collide~\citep{leinhardt12}.  

Taking into account only the accretion condition for embryo-embryo impacts, the final number, mass, orbital elements, and even growth timescale of planets are barely affected~\citep{kokubo10,alexander98}. This is because even though collisions do not lead to accretion, the colliding bodies stay on the colliding orbits after the collision and thus the system is unstable and the next collision occurs shortly.

However, by allowing non-accretionary impacts to both erode the target embryo and to produce debris particles, \cite{chambers13} found that fragmentation does have a noted effect on accretion.  The final stages of accretion are lengthened by the sweep up of collisional fragments.  The planets that formed in simulations with fragmentation had smaller masses and smaller eccentricities than their counterparts in simulations without fragmentation.  

Imperfect accretion also affects the planets' spin rates.  \cite{kokubo10} found that the spin angular momentum of accreted planets was 30\% smaller than in simulations with perfect accretion.  This is because grazing collisions that have high angular momentum are likely to result in a hit-and-run, while nearly head-on collisions that have small angular momentum lead to accretion.  The production of unbound collisional fragments with high angular momentum could further reduce the spin angular velocity.  The effect of non-accretionary impacts on the planetary spins has yet to be carefully studied.  

A final consequence of fragmentation is on the core mass fraction.  Giant impacts lead to an increase in the core mass fraction because the mantle is preferentially lost during imperfect merging events~\citep{benz07,stewart12,genda12}.  However, the sweep-up of these collisional fragments on 100 Myr timescales re-balances the composition of planets to roughly the initial embryo composition~\citep{chambers13}.  We speculate that a net increase in core mass fraction should be retained if the rocky fragments are allowed to collisionally evolve and lose mass.  

\bigskip
\noindent
\textbf{4.5 Effect of outer giant planets}
\bigskip

We now consider the effect of giant planets on terrestrial accretion.  We restrict ourselves to systems with giant planets similar to our own Jupiter and Saturn.  That is, systems with non-migrating giant planets on stable orbits exterior to the terrestrial planet-forming region.  In \S 5.2 we will consider the effects of giant planet migration and planet-planet scattering.  

\begin{figure*}
\epsscale{2}
\plottwo{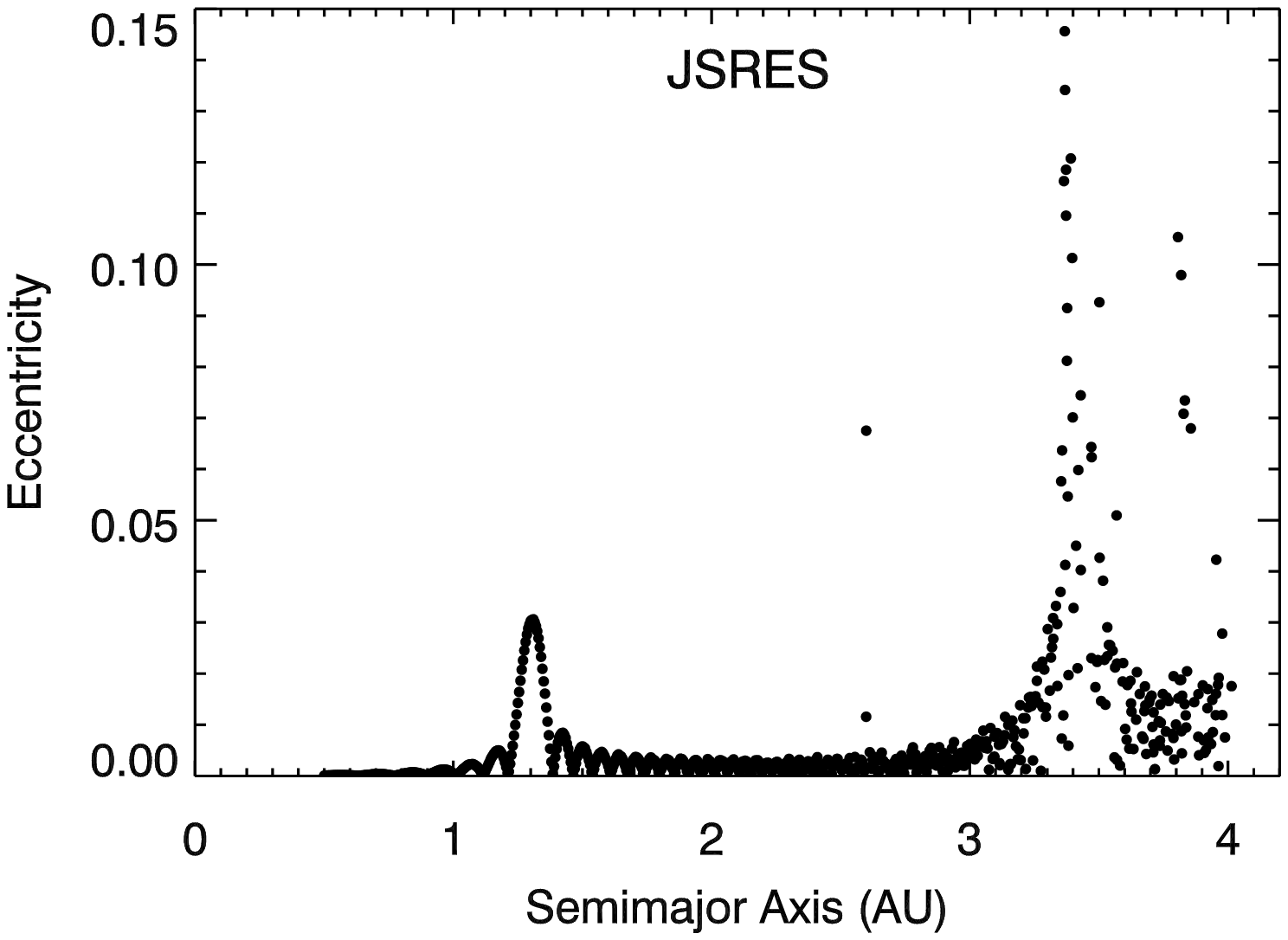}{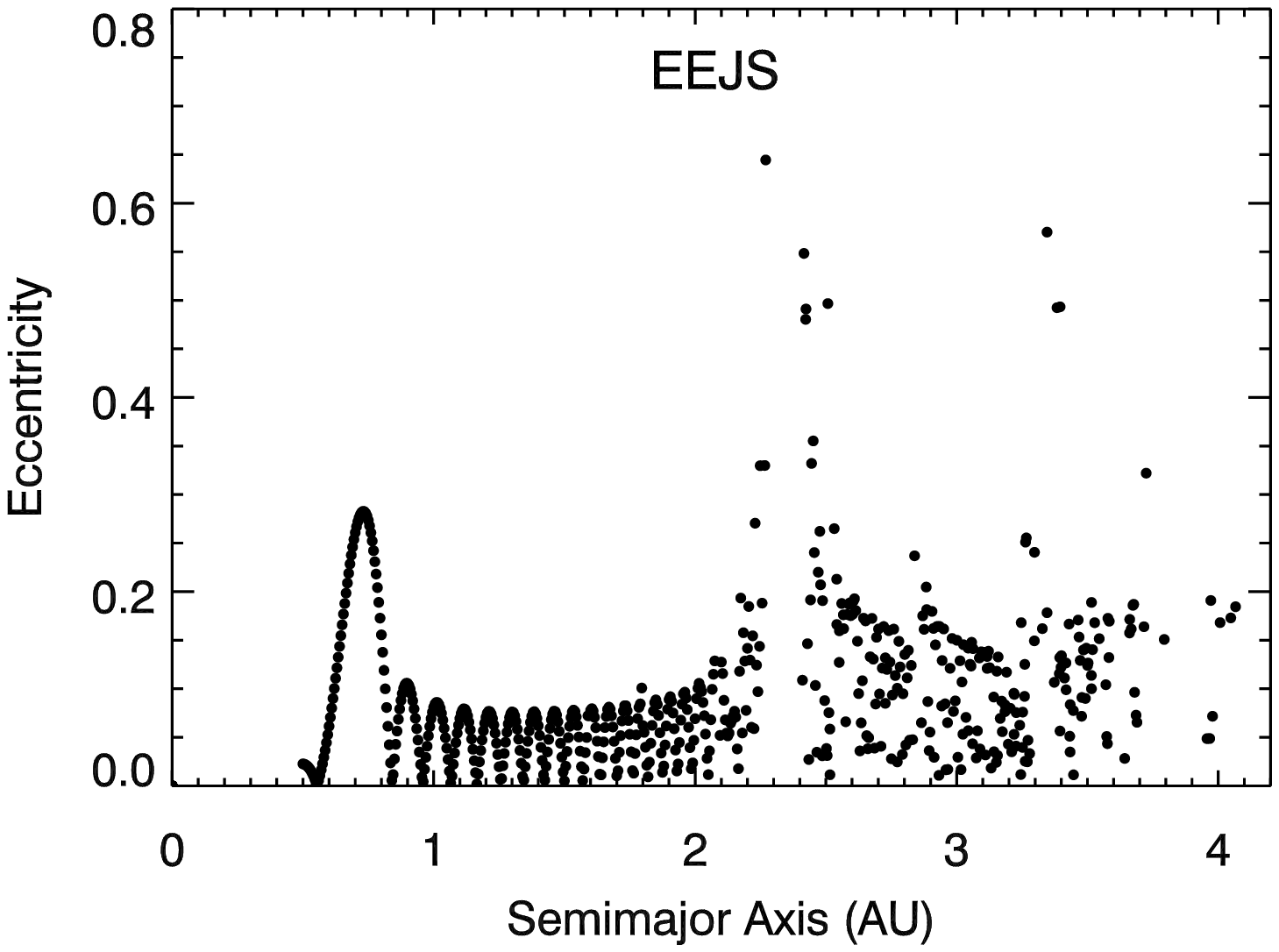}
\caption{\small Excitation of test particles by two configurations of Jupiter and Saturn.  Each panel shows the eccentricities of massless test particles after 1 Myr (giant planets not shown).  Note the difference in the y-axis scale between the two panels.  Each scenario is consistent with the present-day Solar System (see discussion in \S 6).  Jupiter and Saturn are in 3:2 mean motion resonance with semimajor axes of 5.4 and 7.2 AU and low eccentricities in the {\em JSRES} configuration.  The gas giants are at their current semimajor axes of 5.2 and 9.5 AU with eccentricities of 0.1 in the {\em EEJS} configuration.  From \cite{raymond09c}.  }  
\label{fig:testp}
\end{figure*}

The most important effect of giant planets on terrestrial accretion is the excitation of the eccentricities of planetary embryos.  This generally occurs by the giant planet-embryo gravitational eccentricity forcing followed by the transmission of that forcing by embryo-embryo or embryo-planetesimal forcing.  The giant planet forcing typically occurs via mean motion or secular resonances, or secular dynamical forcing.  Giant planet-embryo excitation is particularly sensitive to the giant planets' orbital architecture~\citep{chambers02,levison03,raymond06a}. Figure~\ref{fig:testp} shows the eccentricities of test particles excited for 1 Myr by two different configurations of Jupiter and Saturn~\citep{raymond09c}, both of which are consistent with the present-day Solar System (see \S 6).  The spikes in eccentricity seen in Fig.~\ref{fig:testp} come from specific resonances: in the {\em JSRES} configuration (for ``Jupiter and Saturn in RESonance''), the $\nu_5$ secular resonance at 1.3 AU and the 2:1 mean motion resonance with Jupiter at 3.4 AU; and in the {\em EEJS} configuration (for ``Extra-Eccentric Jupiter and Saturn'') the $\nu_5$ and $\nu_6$ secular resonances at 0.7 and 2.1 AU, and a hint of the 2:1 mean motion resonance with Jupiter at 3.3 AU.  The ``background'' level of excitation seen in Fig.~\ref{fig:testp} comes from secular forcing, following a smooth function of the orbital radius. 

The eccentricity excitation of terrestrial embryos is significant even for modest values of the giant planets' eccentricity.  In Fig.~\ref{fig:testp}, Jupiter and Saturn have eccentricities of 0.01-0.02 in the {\em JSRES} configuration and of 0.1 in the {\em EEJS} configuration.  The test particles in the {\em JSRES} system are barely excited by the giant planets interior to 3 AU; the magnitude of the spike at 1.3 AU is far smaller than the secular forcing anywhere in the {\em EEJS} simulation.  Note also that this figure represents just the first link in the chain.  The eccentricities imparted to embryos are systematically transmitted to the entire embryo swarm, and it is the mean eccentricity of the embryo swarm that dictates the outcome of accretion.  

In a population of embryos with near-circular orbits, the communication zone -- the radial distance across which a given embryo has gravitational contact with its neighbors -- is very narrow.  Embryos grow by collisions with their immediate neighbors.  The planets that form are thus limited in mass by the mass in their immediate vicinity.  In contrast, in a population of embryos with significant eccentricities, the communication zone of embryos is wider.  Each embryo's orbit crosses the orbits of multiple other bodies and, by secular forcing, gravitationally affects the orbits of even more.   This of course does not imply any imminent collisions, but it does mean that the planets that form will sample a wider radial range of the disk than in the case of very low embryo eccentricities.  This naturally produces a smaller number of more massive planets.  Given that collisions preferentially occur at pericenter, the terrestrial planets that form tend to also be located closer-in when the mean embryo eccentricity is larger~\citep{levison03}.  

In systems with one or more giant planets on orbits exterior to the terrestrial planet-forming region, the amplitude of excitation of the eccentricities of terrestrial embryos is larger when the giant planets' orbits are eccentric or closer-in.  The timescale for excitation is shorter when the giant planets are more massive.  Thus, the strongest perturbations come from massive eccentric gas giants.

Simulations have indeed shown that systems with massive or eccentric outer gas giants systematically produce fewer, more massive terrestrial planets~\citep{chambers02,levison03,raymond04}.  However, the efficiency of terrestrial accretion is smaller in the presence of a massive or eccentric gas giant because a fraction of embryos and planetesimals are excited onto orbits that are unstable and are thus removed from the system.  The most common mechanism for the removal of such bodies is by having their eccentricities increased to the point where their orbits cross those of a giant planet, then being ejected entirely from the system into interstellar space.  

The strong outside-in perturbations produced by massive or eccentric outer gas giants also act to accelerate terrestrial planet formation.  This happens for two reasons.  First, when embryos have significant mean eccentricities the typical time between encounters decreases, as long as eccentricities are more strongly perturbed than inclinations.  Second, accretion is slower in the outer parts of planetary systems because of the longer orbital and encounter timescales, and it is these slow-growing regions that are most efficiently cleared by the giant planets' perturbations.  

Given their outside-in influence, outer gas giants also play a key role in water delivery to terrestrial planets.  It should be noted up front that the gas giants' role in water delivery is purely detrimental, at least in the context of outer giant planets on static orbits.  Stimulating the eccentricities of water-rich embryos at a few AU can in theory cause some embryos to be scattered inward and deliver water to the terrestrial planets.  In practice, a much larger fraction of bodies is scattered outward, encounters the giant planets and is ejected from the system than is scattered inward to deliver water~\citep{raymond06b}.  

Finally, simulations with setups similar to the one from Fig.~\ref{fig:sim0} confirm that the presence of one or more giant planets strongly anti-correlates with the water content of the terrestrial planets in those systems~\citep{chambers02,raymond04,raymond06b,raymond07a,raymond09c,obrien06}.  There is a critical orbital radius beyond which a giant planet must lie for terrestrial planets to accrete and survive in a star's liquid water habitable zone~\citep{raymond06a}.  This limit is eccentricity dependent: a zero-eccentricity (single) giant planet must lie beyond 2.5 AU to allow a terrestrial planet to form between 0.8 and 1.5 AU whereas a giant planet with an eccentricity of 0.3 must lie beyond 4.2 AU.  For water to be delivered to the terrestrial planets from a presumed source at 2-4 AU (as in Fig.~\ref{fig:sim0}) the giant planet must be farther still~\citep{raymond06a}.

\section{\textbf{TERRESTRIAL ACCRETION IN EXTRA-SOLAR PLANETARY SYSTEMS}}

Extra-solar planetary systems do not typically look like the Solar System.  To extrapolate to extra-solar planetary systems is therefore not trivial.  Additional mechanisms must be taken into account, in particular orbital migration both of planetary embryos (Type 1 migration) and of gas giant planets (Type 2 migration) and dynamical instabilities in systems of multiple gas giant planets.  

There exists ample evidence that accretion does indeed occur around other stars.  Not only has an abundance of low-mass planets been detected~\citep{mayor11,batalha13}, but the dust produced during terrestrial planet formation~\citep{kenyon04} has also been detected~\citep[e.g.][]{meyer08,lisse08}, including the potential debris from giant embryo-embryo impacts~\citep{lisse09}.  

In this section we first address the issue of the formation of hot Super Earths.  Then we discuss how the dynamics shaping the known systems of giant planets may have sculpted unseen terrestrial planets in those systems.  

\bigskip
\noindent
\textbf{5.1 Hot Super Earths}
\bigskip

Hot Super Earths are extremely common.  Roughly one third to one half of Sun-like (FGK) stars host at least one planet with a mass less than $10 \mearth$ and a period of less than $50-100$ days~\citep{howard10,mayor11}.  The frequency of Hot Super Earths is at least as high around M stars as around FGK stars and possibly higher~\citep{howard12,bonfils13,fressin13}.  Hot Super Earths are typically found in systems of many planets on compact but non-resonant orbits~\citep[e.g.][]{udry07,lovis11,lissauer11}.  


Several mechanisms have been proposed to explain the origin of Hot Super Earths~\citep[see][]{raymond08a}: 1) In situ accretion from massive disks of planetary embryos and planetesimals; 2) Accretion during inward type 1 migration of planetary embryos; 3) Shepherding in interior mean motion resonances with inward-migrating gas giant planets; 4) Shepherding by inward-migrating secular resonances driven by dissipation of the gaseous disk; 5) Circularization of planets on highly-eccentric orbits by star-planet tidal interactions; 6) Photo-evaporation of close-in gas giant planets.

Theoretical and observational constraints effectively rule out mechanisms 3-6.  The shepherding of embryos by migrating resonances (mechanisms 3 and 4) can robustly transport material inward~\citep{zhou05,fogg05,fogg07,raymond06c,mandell07,gaidos07}.  An embryo that finds itself in resonance with a migrating giant planet will have its eccentricity simultaneously excited by the giant planet and damped by tidal interactions with the gaseous disk~\citep{tanaka04,cresswell07}.  As the tidal damping process is non-conservative, the embryo's orbit loses energy and shrinks, removing the embryo from the resonance.  The migrating resonance catches up to the embryo and the process repeats itself, moving the embryo inward, potentially across large distances.  This mechanism is powered by the migration of a strong resonance.  This requires a connection between Hot Super Earths and giant planets.  If a giant planet migrated inward, and the shepherd was a mean motion resonance (likely the 3:2, 2:1 or 3:1 resonance) then hot Super Earths should be found just interior to close-in giant planets, which is not observed.  If a strong secular resonance migrated inward then at least one giant planet on an eccentric orbit must exist exterior to the hot Super Earth, and there should only be a small number of Hot Super Earths.  This is also not observed.  

Tidal circularization of highly-eccentric Hot Super Earths (mechanism 5) is physically possible but requires extreme conditions~\citep{raymond08a}.  Star-planet tidal friction of planets on short-pericenter orbits can rapidly dissipate energy, thereby shrinking and re-circularizing the planets' orbits.  This process has been proposed to explain the origin of hot Jupiters~\citep{ford06,fabrycky07,beauge12}, and the same mechanism could operate for low-mass planets.  Very close pericenter passages -- within 0.02 AU -- are required for significant radial migration~\citep{raymond08a}.  Although such orbits are plausible, another implication of the model is that, given their large prior eccentricities, hot Super Earths should be found in single systems with no other planets nearby.  This is not observed.  

The atmospheres of very close-in giant planets can be removed by photo-evaporation from the host star~\citep[mechanism 6;][]{lammer03,baraffe04,baraffe06,yelle04,erkaev07,hubbard07a,raymond08a,murrayclay09,lopez13}.  The process is driven by UV heating from the central star.  Mass loss is most efficient for planets with low surface gravities extremely close to UV-bright stars.  Within $\sim 0.02$ AU, planets as large as Saturn can be photo-evaporated down to their cores on Gyr timescales.  Since both the photoevaporation rate and the rate of tidal evolution depend on the planet mass, a very close-in rocky planet like {\em Corot-7b}~\citep{leger09} could have started as a Saturn-mass planet on a much wider orbit~\citep{jackson10}.  Although photo-evaporation may cause mass loss in some very close-in planets, it cannot explain the systems of hot Super Earths.  \cite{hubbard07b} showed that the mass distributions of very highly-irradiated planets within 0.07 AU was statistically indistinguishable from the mass distribution of planets at larger distances.  In addition, given the very strong radial dependence of photo-evaporative mass loss, the mechanism is likely to produce systems with a single hot Super Earth as the closest-in planet rather than multiple systems of hot Super Earths.  

Given the current constraints from Kepler and radial velocity observations, mechanisms 1 and 2 -- in situ accretion and type 1 migration -- are the leading candidates to explain the formation of the observed Hot Super Earths.  Of course, we cannot rule out additional mechanisms that have yet to come to light.  

For systems of hot Super Earths to have accreted in situ from massive populations of planetesimals and planetary embryos, their protoplanetary disks must have been very massive~\citep{raymond08a,hansen12,hansen13,chiang13,raymond14}.  The observed systems of hot Super Earths often contain $20-40 \mearth$ in planets within a fraction of an AU of the star~\citep[][]{batalha13}.  Let us put this in the context of simplified power-law disks:  
\begin{equation}
\Sigma = \Sigma_0 \left(\frac{r}{1 {\rm AU}}\right)^{-x}.
\end{equation}
The minimum-mass Solar Nebula (MMSN) model~\citep{weidenschilling77b,hayashi85} has $x = 3/2$, although modified versions have $x = 1/2$~\citep{davis05} and $x \approx 2$~\citep{desch07}.  \cite{chiang13} created a minimum-mass {\em extrasolar} nebula using the Kepler sample of hot Super Earths and found a best fit for $x = 1.6-1.7$ with a mass normalization roughly ten times higher than the MMSN.  However, \cite{raymond14} showed that minimum-mass disks based on Kepler multiple-planet systems actually cover a broad range in surface density slopes and are inconsistent with a universal underlying disk profile.  

Only steep power-law disks allow for a significant amount of mass inside 1 AU.  Consider a disk with a mass of $0.05 M_{\odot}$ extending from zero to 50 AU with an assumed dust-to-gas ratio of 1\%.  This disk contains a total of $150 \mearth$ in solids.  If the disk follows an $r^{-1/2}$ profile (i.e., with $x = 1/2$) then it only contains $0.4 \mearth$ in solids inside 1 AU.  If the disk has $x=1$ then it contains $3 \mearth$ inside 1 AU.  If the disk has $x=1.5-1.7$ then it contains $21-46 \mearth$ inside 1 AU.  Sub-mm observations of cold dust in the outer parts of nearby protoplanetary disks generally find values of $x$ between $1/2$ and 1~\citep{mundy00,looney03,andrews07b}.  However, the inner parts of disks have yet to be adequately measured.


The dynamics of in situ accretion of hot Super Earths would presumably be similar to the well-studied dynamics of accretion presented in sections 3 and 4.  Accretion would proceed faster than at 1 AU due to the shorter relevant timescales, but would consist of embryo-embryo and embryo-planetesimal impacts~\citep{raymond08a}.  However, even if Super Earths accrete modest gaseous envelopes from the disk, these envelopes are expected be lost during the dispersal of the protoplanetary disk under most conditions~\citep{ikoma12}.  This loss process is most efficient at high temperatures, making it hard to explain the large radii of some detected Super Earths.  Nonetheless, Super Earths that form by in situ accretion appear to match several other features of the observed population, including their low mutual inclination orbits and the distributions of eccentricity and orbital spacing~\citep{hansen13}.

\begin{figure}[h]
\epsscale{1}
\plotone{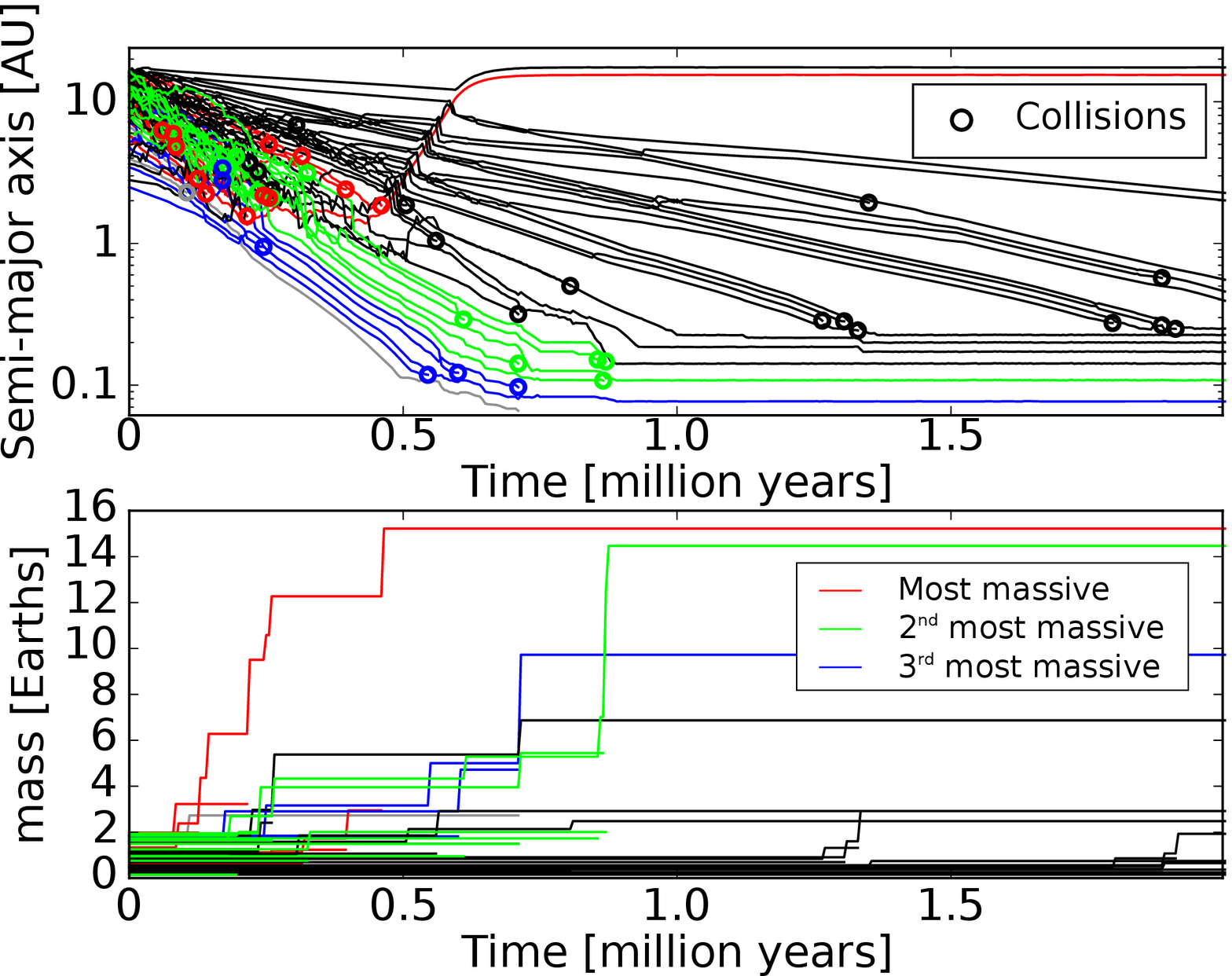}
\plotone{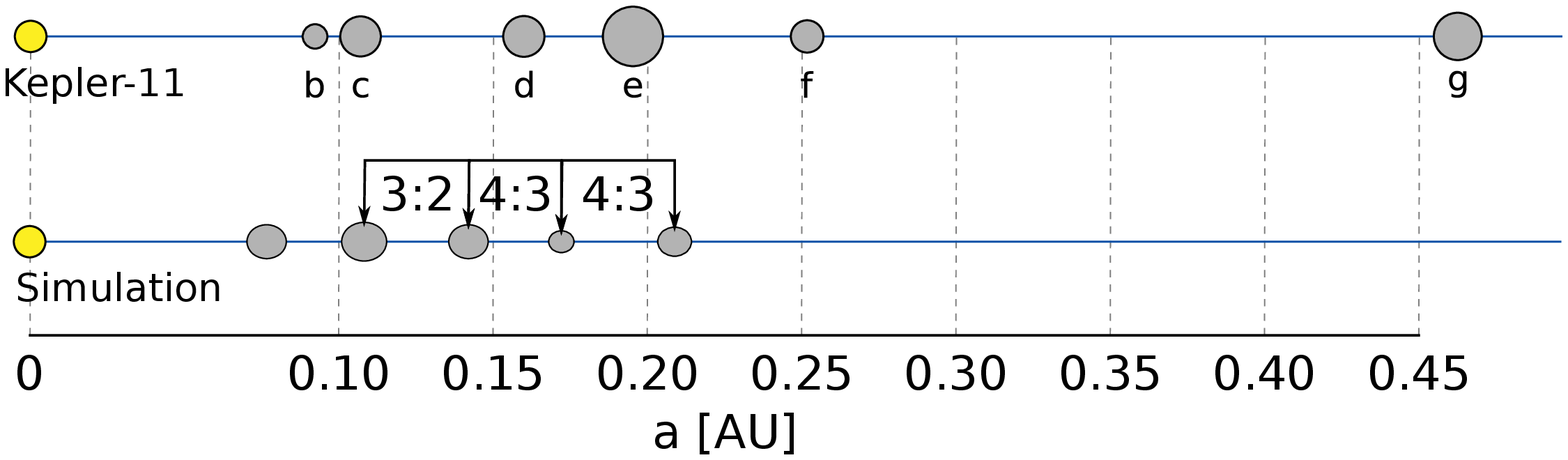}
\caption{\small Formation of a system of hot Super Earths by type 1 migration.  The top panel shows the evolution of the embryos' orbital radii and the bottom panel shows the mass growth.  The red, green and blue curves represent embryos that coagulated into the three most massive planets.  All other bodies are in black.  Only the most massive (red) planet grew large enough to trigger outward migration before crossing into a zone of pure inward migration.  From \cite{cossou13b}.  }  
\label{fig:type1}
\end{figure}

Alternately, the formation of hot Super Earths may involve long-range orbital migration~\citep{terquem07}. Once they reach $\sim 0.1 \mearth$, embryos are susceptible to type 1 migration~\citep{goldreich80,ward86}.  Type 1 migration may be directed inward or outward depending on the local disk properties and the planet mass~\citep{paardekooper11,masset10,kretke12}.  In most disks outward migration is only possible for embryos larger than a few Earth masses.  All embryos therefore migrate inward when they are small.  If they grow quickly enough during the migration then in some regions they can activate their corotation torque and migrate outward.  

A population of inward-migrating embryos naturally forms a resonant chain.  Migration is stopped at the inner edge of the disk~\citep{masset06} and the resonant chain piles up against the edge~\citep{ogihara09}.  If the resonant chain gets too long, cumulative perturbations from the embryos act to destabilize the chain, leading to accretionary collisions and a new shorter resonant chain~\citep{morby08,cresswell08}.  This process can continue throughout the lifetime of the gaseous disk and include multiple generations of inward-migrating embryos or populations of embryos.  

Figure~\ref{fig:type1} shows the formation of a system of hot Super Earths by type 1 migration from \cite{cossou13b}.  In this simulation $60 \mearth$ in embryos with masses of $0.1-2\mearth$ started from 2-15 AU.  The embryos accreted as they migrated inward in successive waves.  One embryo (shown in red in Fig.~\ref{fig:type1}) grew large enough to trigger outward migration and stabilized at a zero-torque zone in the outer disk, presumably to become giant planet core.  The system of hot Super Earths that formed is similar in mass and spacing to the Kepler-11 system~\citep{lissauer11}.  The four outer super Earths are in a resonant chain but the inner one was pushed interior to the inner edge of the gas disk and removed from resonance.  


It was proposed by \cite{raymond08a} that transit measurements of hot Super Earths could differentiate between the in situ accretion and type 1 migration models.  They argued that planets formed in situ should be naked high-density rocks whereas migrated planets are more likely to be dominated by low-density material such as ice.  It has been claimed that planets that accrete in situ can have thick gaseous envelopes and thus inflated radii~\citep{hansen12,chiang13}.  However, detailed atmospheric calculations by \cite{ikoma12} suggest that it is likely that low-mass planets generally lose their atmospheres during disk dispersal.  This is a key point.  If these planets can indeed retain thick atmospheres then simple measurements of the bulk density of Super Earths wold not provide a mechanism for differentiation between the models.  However, if hot Super Earths cannot retain thick atmospheres after forming in situ, then low density planets must have formed at larger orbital distances and migrated inward.  

It is possible that migration and in situ accretion both operate to reproduce the observed hot Super Earths.  The main shortcoming of in situ accretion model is that the requisite inner disk masses are extremely large and do not fit the surface density profiles measured in the outskirts of protoplanetary disks.  Type 1 migration of planetary embryos provides a natural way to concentrate solids in the inner parts of protoplanetary disks.  One can envision a scenario that proceeds as follows.  Embryos start to accrete locally throughout the disk.  Any embryo that grows larger than roughly a Mars mass type 1 migrates inward.  Most embryos migrate all the way to the inner edge of the disk, or at least to the pileup of embryos bordering on the inner edge.  There are frequent close encounters and impacts between embryos.  The embryos form long resonant chains that are successively broken by perturbations from other embryos or by stochastic forcing from disk turbulence~\citep{terquem07,pierens11}.  As the disk dissipates the resonant chain can be broken, leading to a last phase of collisions that effectively mimics the in situ accretion model.  There remains sufficient gas and collisional debris to damp the inclinations of the surviving Super Earths to values small enough to be consistent with observations.  However, that it is possible that many Super Earths actually remain in resonant orbits but with period ratios altered by tidal dissipation~\citep{batygin13}.  


\bigskip
\noindent
\textbf{5.2 Sculpting by giant planets: type 2 migration and dynamical instabilities}
\bigskip

The orbital distribution of giant exoplanets is thought to have been sculpted by two dynamical processes: type 2 migration and planet-planet scattering~\citep{moorhead05,armitage07}.  These processes each involve long-range radial shifts in giant planets' orbits and have strong consequences for terrestrial planet formation in those systems.  In fact, each of these processes has been proposed to explain the origin of hot Jupiters~\citep{lin96,nagasawa08}, so differences in the populations of terrestrial planets, once observed, could help resolve the question of the origin of hot Jupiters.

Only a fraction of planetary systems contain giant planets.  About 14\% of Sun-like stars host a gas giant with period shorter than 1000 days~\citep{mayor11}, although the fraction of stars with more distant giant planets could be significantly higher~\citep{gould10}.  

\begin{figure*}
 \epsscale{0.95}
\plotone{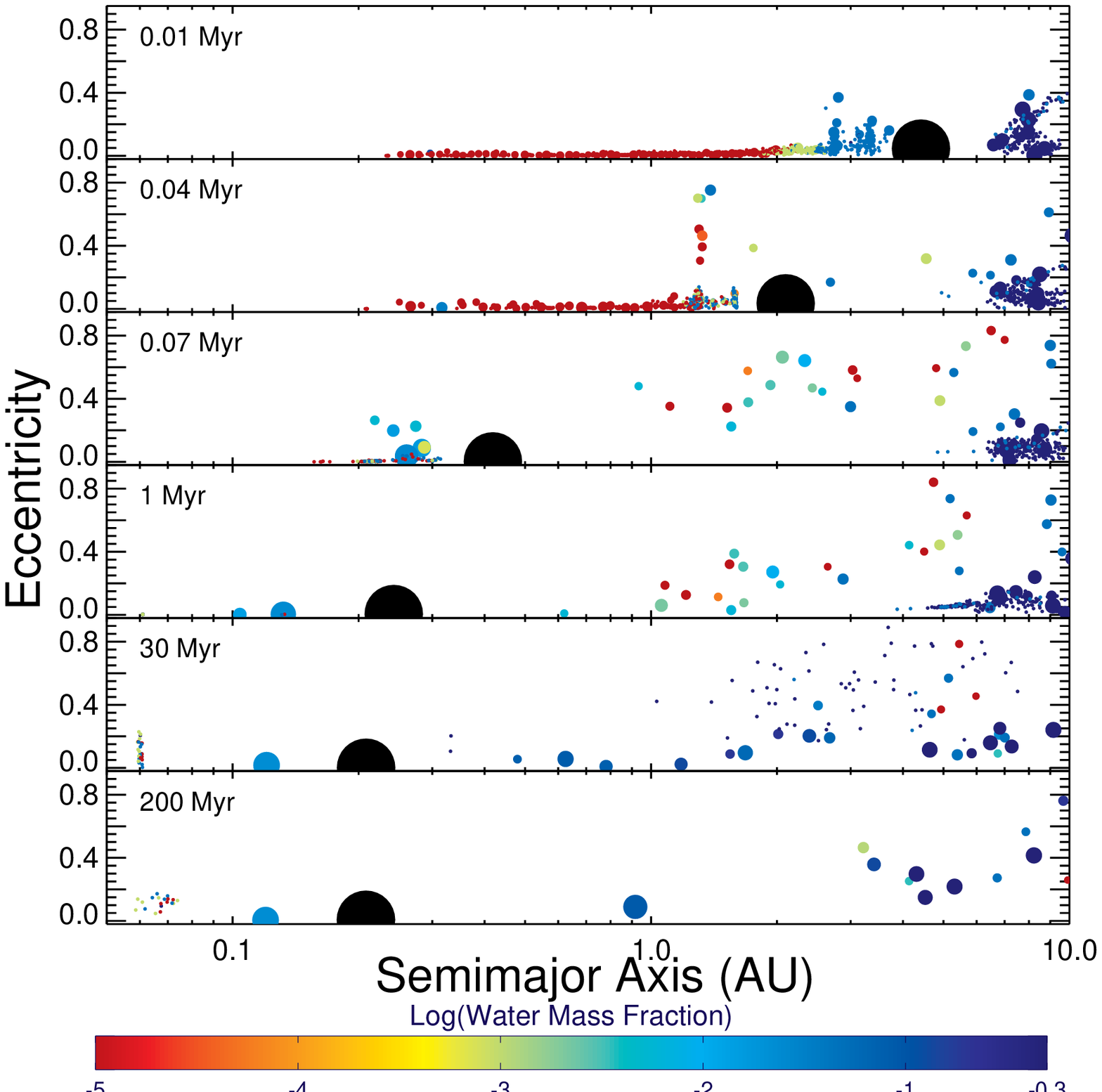}
\plotone{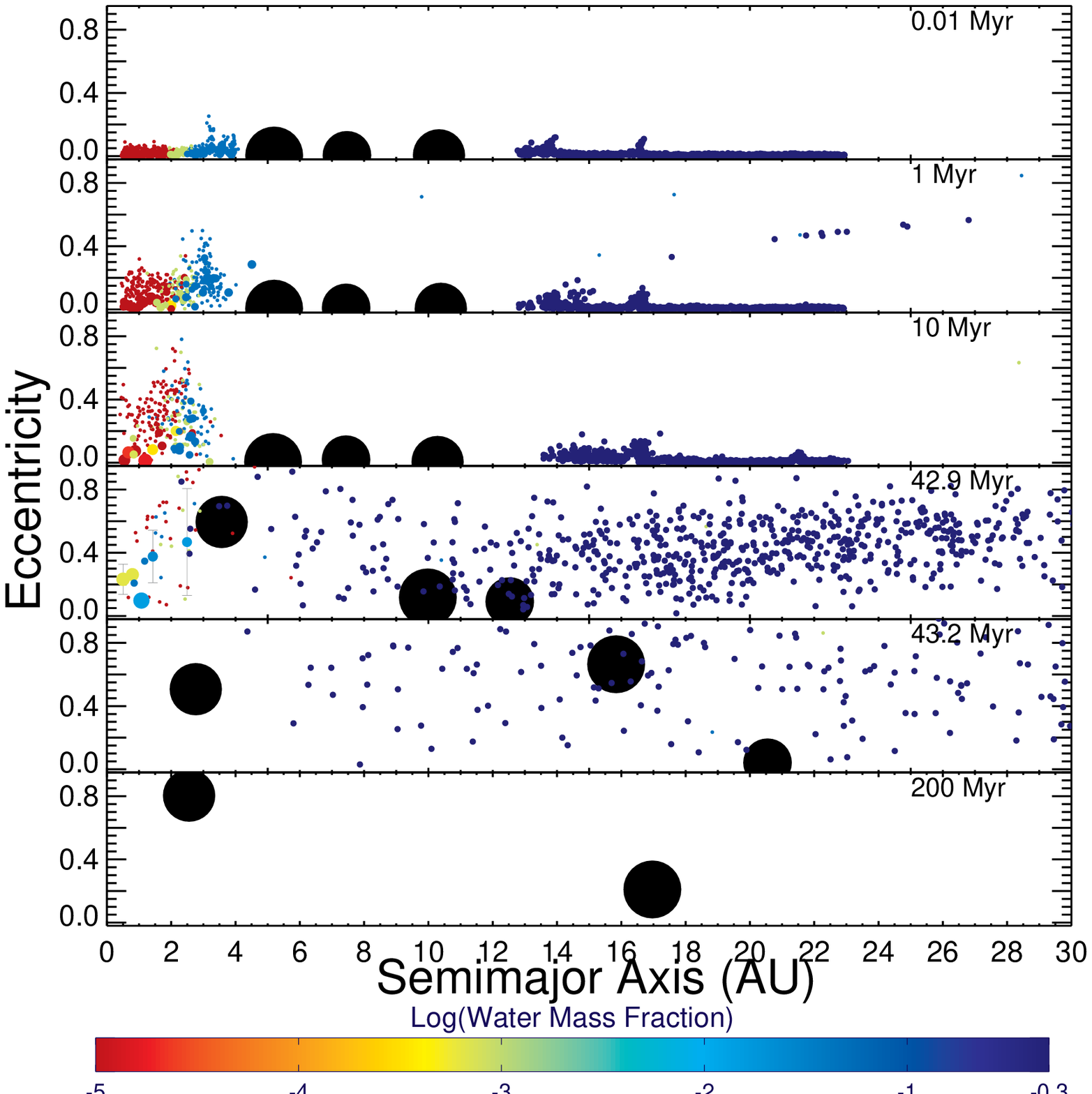}
\caption{\small The effect of giant planet migration (left panel) and dynamical instabilities (right panel) on terrestrial planet formation.  In each panel large black circles represents roughly Jupiter-mass gas giant planets and the smaller circles each represent a planetary embryo or planetesimal.  Colors correspond to water contents (see color bars), and the relative size of each particle (giant planets excepted) refers to their mass$^{1/3}$.  Adapted from simulations by \cite{raymond06c} (left) and \cite{raymond12} (right).}  
\label{fig:aet_giants}
\end{figure*}

When a giant planet becomes massive enough to open a gap in the protoplanetary disk, its orbital evolution becomes linked to the radial viscous evolution of the gas.  This is called Type 2 migration~\citep{lin86,ward97}.  As a giant planet migrates inward it encounters other small bodies in various stages of accretion.  Given the strong damping of eccentricities by the gaseous disk, a significant fraction of the material interior to the giant planet's initial orbit is shepherded inward by strong resonances as explained in \S 5.1~\citep{zhou05,fogg05,fogg07,fogg09,raymond06c,mandell07}.  Indeed, the simulation from the left panel of Figure~\ref{fig:aet_giants} formed two hot Super Earth planets, one just interior to the 2:1 and 3:1 resonance.  The orbits of the two planets became destabilized after several Myr, collided and fused into a single $4 \mearth$ hot Super Earth.  There also exists a population of very close-in planetesimals in the simulation from Fig.~\ref{fig:aet_giants}; these were produced by the same shepherding mechanism as the hot Super Earths but, because the dissipative forces from gas drag were so much stronger for these objects than the damping due to disk-planet tidal interactions felt by the embryos~\citep{adachi76,ida08b}, they were shepherded by a much higher-order resonance, here the 8:1.

Planetesimals or embryos that come too close to the migrating giant are scattered outward onto eccentric orbits.  These orbits are slowly re-circularized by gas drag and dynamical friction.  On 10-100 Myr or longer timescales a second generation of terrestrial planets can form from this scattered material~\citep{raymond06c,mandell07}.  The building blocks of this new generation of planets are significantly different than the original ones.  This new distribution is comprised of two components: bodies that originated across the inner planetary system that were scattered outward by the migrating gas giant, and bodies that originated exterior to the gas giant.  When taking into account the original location of these protoplanets, the effective feeding zone of the new terrestrial planets essentially spans the entire planetary system.  This new generation of terrestrial planets therefore inevitably contains material that condensed at a wide range of orbital distances.  Their volatile contents are huge.  Indeed, the water content of the $3 \mearth$ planet that formed at 0.9 AU (in the shaded habitable zone) in Fig.~\ref{fig:aet_giants} is roughly 10\% by mass.  Even if 90\% of the water were lost during accretion, that still corresponds to ten times Earth's water content (by mass), meaning that this planet is likely to be covered in global oceans.  

The simulation from Fig.~\ref{fig:aet_giants} showed the simple case of a single giant planet on a low-eccentricity ($e \approx 0.05$) migrating through a disk of growing planetesimals and embryos.  Migration would be more destructive to planet formation under certain circumstances.  For example, if migration occurs very late in the evolution of the disk then less gas remains to damp the eccentricities of scattered bodies.  This is probably more of an issue for the formation of hot Super Earths than for scattered embryos: since the viscous timescale is shorter closer-in, much of the inner disk may in fact drain onto the star during type 2 migration~\citep{thommes08} and reduce the efficiency of the shepherding mechanism.  In addition, multiple giant planets may often migrate inward together.  In that case the giant planets' eccentricities would likely be excited to modest values, and any embryo scattered outward would likely encounter another giant planet, increasing the probability of very strong scattering events onto unbound orbits.  

Although type 2 migration certainly does not provide a comfortable environment for terrestrial accretion, planet-planet scattering is far more disruptive.  The broad eccentricity distribution of observed giant exoplanets is naturally reproduced if at least 75\% of the observed planets are the survivors of violent dynamical instabilities~\citep{chatterjee08,juric08,raymond10}.  It is thought that giant planets form in multiple systems on near-circular orbits but in time, perturbations destabilize these systems and lead to a phase of close gravitational encounters.  Repeated planet-planet scattering usually leads to the ejection of one or more giant planets~\citep[][; see chapter by Davies et al]{rasio96,weidenschilling96}.  The large eccentricities of the observed planets are essentially the scars of past instabilities.

Instabilities are also destructive for terrestrial planets or their building blocks.  The timing of instabilities is poorly-constrained, although it is thought that many instabilities may be triggered by either migration in systems of multiple gas giants~\citep[][]{adams03,moorhead05} or by the removal of damping during the dissipation of the gaseous disk~\citep{moeckel08,matsumura10,moeckel12}.  On the other hand, systems of planets on more widely-spaced orbits or systems with wide binary companions may naturally experience instabilities on Gyr timescales~\citep{marzari02,kaib13}.  Although early instabilities may allow for additional sources of damping via gas drag from remaining gas and dynamical friction from abundant planetesimals, in practice the timing of the instability makes little difference for the survival of terrestrial bodies~\citep{raymond12}.

Instabilities between Jupiter-sized planets typically only last for $\sim10^5$ years.  When a giant planet is scattered onto a highly-eccentric orbit, even if it only lasts for a relatively short time, very strong secular forcing can drive the orbits of inner terrestrial bodies to very high eccentricities.  The outcome of the perturbation is extremely sensitive to the proximity of the giant planet to the terrestrial planet zone: giant planets whose pericenter distances come within a given separation act so strongly that terrestrial planets or embryos are driven entirely into the central star~\citep{veras05,veras06,raymond11,raymond12}.  The giant planet instabilities that are the least disruptive to the terrestrial planets are those that are very short in duration, that are confined to the outer parts of the planetary system, or that result in a collision between giant planets.  

The right panel of Figure~\ref{fig:aet_giants} shows a simulation in which all terrestrial bodies were removed from the system by an instability between three $\sim$Jupiter-mass giant planets that occurred after 42 Myr.  During the first 42 Myr of the simulation, accretion in the inner disk proceeded in the same manner as in Fig.~\ref{fig:sim0}.  Once the instability was triggered after 42.8 Myr, the inner disk of planets -- including two planets that had grown to nearly an Earth mass -- were driven into the central star.  The entire outer disk of planetesimals was ejected by repeated giant planet-planetesimal scattering over the next few Myr~\citep{raymond12}.  

Instabilities systematically perturb both the terrestrial planet-forming region and outer disks of planetesimals.  The dynamics of gas giant planets thus creates a natural correlation between terrestrial planets and outer planetesimal disks.  On Gyr timescales planetesimal disks collisionally grind down and produce cold dust that is observable at wavelengths as debris disks~\citep{wyatt08,krivov10}.  On dynamical grounds, \cite{raymond11,raymond12} predicted a correlation between debris disks and systems of low-mass planets, as each of these forms naturally in dynamically calm environments, i.e. in systems with giant planets on stable orbits or in systems with no gas giants.

\section{\textbf{FORMATION OF THE SOLAR SYSTEM'S TERRESTRIAL PLANETS}}

A longstanding goal of planet formation studies has been to reproduce the Solar System using numerical simulations.  Although that goal has not yet been achieved, substantial progress has been made. 

Jupiter and Saturn are key players in this story.  Their large masses help shape the final stages of terrestrial accretion (\S 4.5). However, there exist few constraints on their orbits during late-stage terrestrial accretion, and these are model-dependent.  

The Nice model~\citep[e.g.,][]{tsiganis05,morby07} proposes that the {\em Late Heavy Bombardment} (LHB) -- a spike in the impact rate on multiple Solar System bodies that lasted from roughly 400 until 700 Myr after the start of planet formation~\citep{tera74,cohen00,chapman07} -- was triggered by an instability in the giant planets' orbits.  The instability was triggered by gravitational interactions between the giant planets and a disk of planetesimals exterior to the planets' orbits comprising perhaps $30-50\mearth$.  Before the Nice model instability,  the giant planets' orbits would have been in a more compact configuration, with Jupiter and Saturn interior to the 2:1 resonance and perhaps lodged in 3:2 resonance.  Although there is no direct constraint, hydrodynamical simulations indicate that the gas giants' eccentricities were likely lower than their current values, probably around 0.01-0.02~\citep{morby07}. 

An alternate but still self-consistent assumption is that the gas giants were already at their current orbital radii during terrestrial accretion.  In that case, Jupiter and Saturn must have had slightly higher eccentricities than their current ones because scattering of embryos during accretion tends to modestly decrease eccentricities~\citep[e.g.][]{chambers02}.  In this scenario, an alternate explanation for the LHB is needed.  

In this section we first consider ``classical'' models that assume that the orbits of the giant planets were stationary (\S 6.1).  Based on the above arguments we consider two reasonable cases.  In the first case, Jupiter and Saturn were trapped in 3:2 mean motion resonance at 5.4 and 7.2 AU with low eccentricities ($e_{giants} \approx 0.01-0.02$).  In the second, Jupiter and Saturn were at their current orbital radii but with higher eccentricities ($e_{giants} = 0.07-0.1$).  

Of course, Jupiter and Saturn's orbits need not have been stationary at this time.  It is well-known that giant planets' orbits can migrate long distances, inward or outward, driven by exchanges with the gaseous protoplanetary disk~\citep[e.g.][]{lin86,veras04} or a disk of planetesimals~\citep[e.g.][]{fernandez84,murray98}.  Although the last phases of accretion are constrained by Hf-W measurements of Earth samples to occur after the dissipation of the typical gas disk, giant planet migration at early times can sculpt the population of embryos and thus affect the ``initial conditions'' for late-stage growth.  

\begin{deluxetable}{l|ccccccl}
\footnotesize
\tablewidth{0pt}
\tablecaption{Success of different models in matching inner Solar System constraints\tablenotemark{1}}
\renewcommand{\arraystretch}{.6}
\tablehead{
\\
\colhead{Model} &  
\colhead{AMD} &
\colhead{RMC} &
\colhead{$M_{Mars}$} &
\colhead{$T_{form}$} &
\colhead{Ast. Belt} &
\colhead{$WMF_\oplus$} & 
\colhead{Comments}
}
\startdata
Resonant Jup, Sat & $\checkmark$ & $\times$ &$\times$ & $\checkmark $ & $\times$ &$\checkmark$ & Consistent with Nice model
\\
Eccentric Jup, Sat & $\sim$ & $\sim$ &$\checkmark$ & $\checkmark $ & $\checkmark$ &$\times$ & Not consistent with Nice model
\\

Grand Tack & $\checkmark$ &$\checkmark$ &$\checkmark$ & $\sim$ & $\checkmark$ &$\checkmark$ & Requires tack at 1.5 AU
\\
Planetesimal-driven & $\checkmark$ & $\times$ & $\times$ & $\checkmark$ & $\times$ &$\checkmark$ & Requires other source of LHB \\
migration
\\
\enddata
\vskip -0.5cm
\tablenotetext{1}{\footnotesize A check (``$\checkmark$'') represents success in reproducing a given constraint, a cross (``$\times$'') represents a failure to reproduce the constraint, and a twiddle sign (``$\sim$'') represents a ``maybe'', meaning success in reproducing the constraints in a fraction of cases.  The constraints are, in order, the terrestrial planets' angular momentum deficit $AMD$ and radial mass concentration $RMC$ (see also Fig.~\ref{fig:amdrmc}), Mars' mass, Earth's formation timescale, the large-scale structure of the asteroid belt, and the delivery of water to Earth (represented by Earth's water mass fraction $WMF_\oplus$). }
\normalsize
\end{deluxetable}

While the Nice model relies on a delayed planetesimal-driven instability, earlier planetesimal-driven migration of the giant planets has recently been invoked~\citep{agnor12}.  In \S 6.2 we consider the effect of this migration, which must occur on a timescale shorter than Mars' measured few Myr accretion time~\citep{dauphas11} to have an effect.  Finally, in \S 6.3 we describe a new model called the {\em Grand Tack}~\citep{walsh11} that invokes early gas-driven migration of Jupiter and Saturn.  

It is possible that disks are not radially smooth, or at least that planetesimals do not form in a radially-uniform way~\citep[e.g.][]{johansen07,chambers10}.  \cite{jin08} proposed that a discontinuity in viscosity regimes at $\sim$2 AU could decrease the local surface density and thus form a small Mars.  However, the dip produced is too narrow to cut off Mars' accretion~\citep{raymond09c}.  It has also been known for decades that an embryo distribution with an abrupt radial edge naturally forms large planets within the disk but small planets beyond the edge~\citep{wetherill78}.  This ``edge effect'' can explain the large Earth/Mars mass ratio (see below). 

Table 2 summarizes the ability of various models to reproduce the observational constraints discussed in \S 2.  

\bigskip
\noindent
\textbf{6.1 Classical models with stationary gas giants}
\bigskip


Fig.~\ref{fig:testp} shows how the giant planets excite the eccentricities of test particles for each assumption~\citep{raymond09c}.  In the left panel (labeled JSRES for ``Jupiter and Saturn in RESonance'') the giant planets are in a low-eccentricity compact configuration consistent with the Nice model whereas in the right panel (labeled EEJS for ``Extra-Eccentric Jupiter and Saturn'') the giant planets have significant eccentricities and are located at their current orbital radii.  The much stronger eccentricity excitation imparted by eccentric gas giants and the presence of strong resonances such as the $\nu_6$ resonance seen at 2.1 AU in the right panel of Fig.~\ref{fig:testp} have a direct influence on terrestrial planet formation.  



Simulations with Jupiter and Saturn on circular orbits reproduce several aspects of the terrestrial planets~\citep{wetherill78,wetherill96,wetherill85,chambers98,morby00,chambers01,raymond04,raymond06b,raymond07a,raymond09c,obrien06,morishima10}.  Simulations typically form about the right number (3-5) of terrestrial planets with masses comparable to their actual masses.  Earth analogs tend to complete their accretion on 50-100 Myr timescales, consistent with geochemical constraints. Simulations include late giant impacts between embryos with similar characteristics to the one that is thought to have formed the Moon~\citep{cuk12,canup12}.  Embryos originating at 2.5-4 AU, presumed to be represented by carbonaceous chondrites and therefore to be volatile-rich, naturally deliver water to Earth during accretion (see Fig.~\ref{fig:sim0}).  

There are three problems.  First and most importantly, simulations with Jupiter and Saturn on circular orbits are unable to form good Mars analogs.  Rather, planets at Mars' orbital distance are an order of magnitude too massive, a situation called the {\em small Mars problem}~\citep{wetherill91,raymond09c}.  Second, the terrestrial planet systems that form tend to be far too spread out radially.  Their radial mass concentration $RMC$ (see Eq. 2) are far smaller than the Solar System's value of 89.9 (see Table 1).  Third, large ($\sim$Mars-sized) embryos are often stranded in the asteroid belt.  All three of these problems are related: the large $RMC$ in these systems is a consequence of too much mass existing beyond 1 AU.  This mass is in the form of large Mars analogs and embryos in the asteroid belt.  

Simulations starting with Jupiter and Saturn at their current orbital radii but with larger initial eccentricities ($e = 0.07-0.1$) reproduce many of the same terrestrial planet constraints~\citep{raymond09c,morishima10}. Simulations tend to again form the same number of terrestrial planets with masses comparable to the actual planets'.  Moon-forming impacts also occur.  Beyond this the accreted planets contrast with those that accrete in simulations with circular gas giants.  With eccentric Jupiter and Saturn, the terrestrial planets accrete faster, in modest agreement with Earth's geochemical constraints.  The delivery of water to Earth is much less efficient.  But Mars analogs form with about the right mass!

\begin{figure}[h]
 \includegraphics[width=0.45\textwidth]{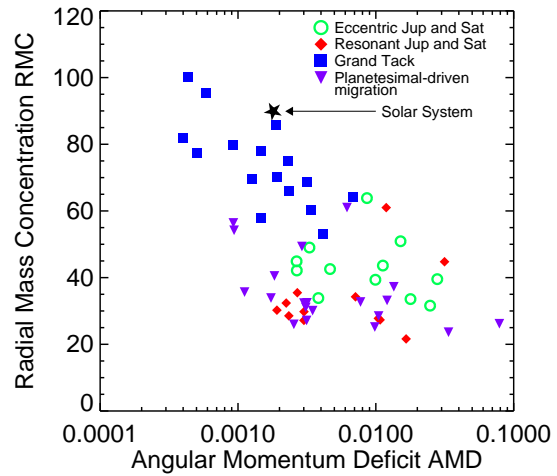}
 \caption{Orbital statistics of the terrestrial planet systems formed in different models.  The configuration of each system is represented by its angular momentum deficit and radial mass concentration values; see section 2.1 for the definition of these terms.  The simulations with eccentric and resonant gas giants are from \cite{raymond09c}, those including planetesimal-driven migration of the gas giants are from \cite{lykawka13}, and the Grand Tack simulations are from \cite{obrien13}. }
\label{fig:amdrmc}
\end{figure}

In these simulations, a strong secular resonance with Saturn -- the $\nu_6$ at 2.1 AU -- acts to clear out the material in the inner asteroid belt and in Mars' vicinity.  The resonance is so strong that bodies that are injected into it are driven to very high eccentricities and collide with the Sun within a few Myr~\citep{gladman97}.  Any embryo from the inner planetary system that is scattered out near the $\nu_6$ is quickly removed from the system.  The Mars region is quickly drained and a small Mars forms.  The $\nu_6$ acts as a firm outer edge such that the terrestrial planet systems form in more compact configurations, with $RMC$ values that approach the Solar System's (but still remain roughly a factor of two too small; see Fig.\ref{fig:amdrmc}).  The $AMD$ of the terrestrial planets are systematically higher than the Solar System value because the planetesimals that could provide damping at late times are too efficiently depleted.  The terrestrial planet forming region is effectively cut off from the asteroid belt by the resonance, and water delivery is inefficient.  If the gravitational potential from the dissipating gas disk is accounted for, the $\nu_5$ and $\nu_6$ resonances sweep inward and can perhaps shepherd water-rich embryos in to Earth's feeding zone by the same mechanism presented in Sec 5.2~\citep{thommes08,morishima10}.  However, hydrodynamical simulations suggest that Jupiter and Saturn's eccentricities are unlikely to remain high enough during the gaseous disk phase for this to occur~\citep[e.g.][]{morby07,pierens11}.  

The early orbits of Jupiter and Saturn sculpt dramatically different terrestrial planet systems.  Systems with gas giants on circular orbits form Mars analogs that are far too large and strand embryos in the asteroid belt.  Systems with gas giants on eccentric orbits do not deliver water to Earth and have eccentricities that are too large.   To date, no other configuration of Jupiter and Saturn with static orbits has been shown to satisfy all constraints simultaneously.  

To quantify the failings of the classical model, Figure~\ref{fig:amdrmc} shows the angular momentum deficit $AMD$ and radial mass concentration $RMC$ statistics for simulated terrestrial planets under the two assumptions considered here.  The accreted planets are far too radially spread out (have small $RMC$ values).  In many cases their orbits are also too excited, with larger $AMD$ values than the actual terrestrial planets'.  


\bigskip
\noindent
\textbf{6.2 Accretion with planetesimal-driven migration of Jupiter and Saturn}
\bigskip

If Jupiter and Saturn formed in a more compact orbital configuration, then the migration to their current configuration may have perturbed the terrestrial planets, or even the building blocks of the terrestrial planets if their formation was not complete.  \cite{brasser09,brasser13} and \cite{agnor12} simulated the influence of planetesimal-driven migration of the giant planets on the terrestrial planets assuming that the migration occurred late, after the terrestrial planets were fully-formed. They found that if Jupiter and Saturn migrated with eccentricities comparable to their present-day values, a smooth migration with an exponential timescale characteristic of planetesimal-driven migration ($\tau \sim$ 5-10 Myr) would have perturbed the eccentricities of the terrestrial planets to values far in excess of the observed ones.  To resolve this issue, \cite{brasser09,brasser13} suggested a Òjumping JupiterÓ in which encounters between an ice giant and Jupiter caused Jupiter and Saturn's orbits to spread much faster than if migration were driven solely by encounters with planetesimals~\citep[see also][]{morby10}.  On the other hand, \cite{agnor12} suggested that the bulk of any giant planet migration occurred during accretion of terrestrial planets.  

Whenever the migration occurred, the degree of eccentricity excitation of Jupiter and Saturn is constrained by the dynamics of resonance crossing.  Jupiter and Saturn are naturally excited to $e_{giants} \sim 0.05$ but cannot reach the higher eccentricities invoked by the eccentric Jupiter and Saturn model described above~\citep{tsiganis05}.  Given that the eccentricity excitation is the key difference between this model and those with stationary giant planets discussed above, the only free parameter is the timing of the eccentricity excitation.  

Two recent papers simulated the effect of planetesimal-driven migration of Jupiter and Saturn's orbits on terrestrial planet formation~\citep{walsh11b,lykawka13}.  In both studies terrestrial planets accrete from a disk of material which stretches from $\sim$0.5 AU to 4.0 AU. In \cite{walsh11b}, Jupiter and Saturn are initially at 5.4 and 8.7 AU respectively (slightly outside the 2:1 mean motion resonance), with eccentricities comparable to the current ones, and migrate to 5.2 and 9.4 AU with an e-folding time of 5 Myr.  In their simulations Mars is typically far too massive and the distribution of surviving planetesimals in the asteroid belt is inconsistent with the observed distribution. \cite{lykawka13} performed similar simulations but included the 2:1 resonance crossing of Jupiter and Saturn, which provides a sharp increase in the giant planets' eccentricities and thus in the perturbations felt by the terrestrial planets.  They tested the timing of the giant planets' 2:1 resonance crossing between 1 and 50 Myr.  They found the expected strong excitation in the asteroid belt once the giant planets' eccentricities increased, but the perturbations were too small to produce a small Mars.  Although they produced four Mars analogs in their simulations, they remained significantly more massive than the real Mars, accreted on far longer timescales than the geochemically-constrained one, and stranded large embryos in the asteroid belt.  Their $AMD$ and $RMC$ values remain incompatible with the real Solar System (Fig.~\ref{fig:amdrmc}). 

If another mechanism is invoked to explain the late heavy bombardment, planetesimal-driven migration of Jupiter and Saturn is plausible.  However, it does not appear likely to have occurred as it is incapable of solving the Mars problem.  

\bigskip
\noindent
\textbf{6.3 The Grand Tack model}
\bigskip

Prior to 2009, several studies of terrestrial accretion had demonstrated an edge effect in terrestrial accretion.  A distribution of embryos with an abrupt edge naturally produces a large mass gradient between the massive planets that formed within the disk and the smaller planets that were scattered beyond the disk's edge~\citep{wetherill78,wetherill91,chambers98,agnor99,chambers01,kominami04}.  These studies had outer edges at 1.5-2 AU and generally considered their initial conditions a deficiency imposed by limited computational resources.  

\cite{hansen09} turned the tables by proposing that, rather than a deficiency, initial conditions with edges might actually represent the true initial state of the disk.  Indeed, \cite{morishima08} and \cite{hansen09} showed that most observed constraints could be reproduced by a disk of embryos spanning only from 0.7 to 1~AU.  Earth and Venus are massive because they formed within the annulus whereas Mars and Mercury's small masses are explained as edge effects, embryos that were scattered exterior and interior, respectively, to the annulus at early times, stranding and starving them.  Mars analogs consistently accrete on the short observed timescale. The main unanswered question in these studies was the origin of the edges of the annulus. 

 
 \begin{figure}[h]
 \epsscale{0.95}
\plotone{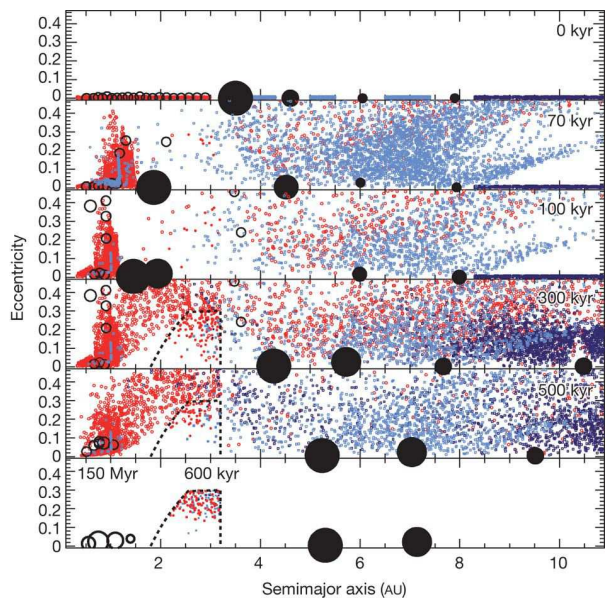}
\caption{\small Evolution of the Grand Tack model~\citep{walsh11}.  The large black dots represent the four giant planets, with sizes that correspond to their approximate masses. Red symbols indicate S-class bodies and blue ones C-class bodies.  There exist two categories of C-class objects that originate between and beyond the giant planets' orbits.  Open circles indicate planetary embryos.  The evolution of the particles includes drag forces imparted by an evolving gaseous disk.}  
\label{fig:aet_GT}
\end{figure}

\cite{walsh11} presented a mechanism to produce the outer edge of the disk by invoking migration of the giant planets to dramatically sculpt the distribution of solid material in the inner Solar System.  Given that gas giant planets must form in the presence of gaseous disks and that these disks invariably drive radial migration~\citep{ward97}, it is natural to presume that Jupiter and Saturn must have migrated to some extent.  A Jupiter-mass planet naturally carves an annular gap in the gaseous disk and migrates inward on the local viscous timescale~\citep{lin86}.  In contrast, a Saturn-mass planet migrates much more quickly because of a strong gravitational feedback during disk clearing~\citep{masset03}.  Assuming that Jupiter underwent rapid gas accretion before Saturn, hydrodynamical simulations show that Jupiter would have migrated inward relatively slowly.  When Saturn underwent rapid gas accreted it migrated inward quickly, caught up to Jupiter and became trapped in 2:3 resonance.  At this point the direction of migration was reversed and Jupiter ``tacked'', that is it changed its direction of migration~\citep{masset01,morby07,pierens08,pierens11,dangelo12}.  The outward migration of the two gas giants slowed as the gaseous disk dissipated, stranding Jupiter and Saturn on resonant orbits.  This naturally produces the initial conditions for a recently revised version of the Nice model~\citep{morby07,levison11}, with Jupiter at 5.4 AU and Saturn at 7.2 AU.  

This model is called the {\em Grand Tack}.  One cannot know the precise migration history of the gas giants {\it a priori} given uncertainties in disk properties and evolution.  \cite{walsh11} anchored Jupiter's migration reversal point at 1.5~AU because this truncates the inner disk of embryos and planetesimals at 1.0~AU, creating an outer edge at the same location as invoked by \cite{hansen09}.  Jupiter's formation zone was assumed to be $\sim3-5$~AU~\citep[although a range of values was tested by][]{walsh11}, in the vicinity of the snow line~\citep[e.g.][]{sasselov00,kornet04,martin12}, presumably a favorable location for giant planet formation.  The Grand Tack model also proposes that the compositional gradient seen in the asteroid belt can be explained by the planetesimals' formation zones.  Volatile-poor bodies (``S-class") are primarily located in the inner belt and volatile-rich bodies (``C-class") primarily in the outer belt~\citep{gradie82,demeo13}. The Grand Tack scenario presumes that S-class bodies formed interior to Jupiter's initial orbit and that C-class bodies formed exterior.  

The evolution of the Grand Tack is illustrated in Figure~\ref{fig:aet_GT}~\citep{walsh11}.  Jupiter and Saturn's inward migration scattered S-class planetesimals from the inner disk, with $\sim$10\% ending on eccentric orbits beyond the giant planets. Meanwhile a large fraction of planetesimals and embryos were shepherded inward by the same mechanism discussed in \S 5.2 onto orbits inside 1 AU.  Following Jupiter's ``tack'' the outward-migrating gas giants first encountered the scattered S-class planetesimals, about 1\% of which were scattered inward onto stable orbits in the asteroid belt. The giant planets then encountered the disk of C-class planetesimals that originated beyond Jupiter's orbit.  Again, a small fraction ($\sim1\%$) were scattered inward and trapped in the asteroid belt.  The final position of a scattered body depends on the orbital radius of the scattering body, in this case Jupiter.  Jupiter was closer in when it scattered the S-class planetesimals and farther out when it scattered the C-class planetesimals.  The S-class bodies were therefore preferentially implanted in the inner part of the asteroid belt and the C-class bodies preferentially in the outer part of the belt, as in the present-day belt~\citep{gradie82,demeo13}.  The total mass of the asteroid population is set by the need to have $\sim 2 \mearth$ of material remaining in the inner truncated disk of embryos and planetesimals (to form the planets).  This requirement for the planets sets the total mass in S-class bodies implanted into the asteroid belt as they originate from the same inner disk. The current ratio of S-class to C-class asteroids sets the mass in outer disk planetesimals.  


The Grand Tack model reproduces many aspects of the terrestrial planets. Planets that accrete from a truncated disk have similar properties to those in \cite{hansen09} and \cite{morishima08}. Earth/Mars mass ratios are close matches to the actual planets, and Mars' accretion timescale is a good match to Hf/W constraints.  Figure~\ref{fig:amdrmc} shows that the angular momentum deficit $AMD$ is systematically lower than in simulations of the classical model (\S 6.1) and the radial mass concentration $RMC$ is systematically higher~\citep{walsh11,obrien13}.  In contrast with other models, the Grand Tack simulations provide a reasonable match to the inner Solar System.  

The Grand Tack delivers water-rich material to the terrestrial planets by a novel mechanism.  As Jupiter and Saturn migrate outward, they scatter about 1\% of the C-class asteroids that they encountered onto stable orbits in the asteroid belt.  And for every implanted C-type asteroid, 10-20 C-class bodies are scattered onto {\em unstable} orbits that cross the orbits of the terrestrial planets.  These scattered C-class planetesimals accrete with the growing terrestrial planets and naturally deliver water.  The amount of water-rich material accreted by Earth is less than in classical simulations with stationary giant planets like the one presented in Fig.~\ref{fig:sim0}, but is still significantly larger than the Earth's current water budget~\citep{obrien13}.  The chemical signature of the delivered water is the same as C-type asteroids (and therefore carbonaceous chondrites), and thus provides a match to the signature of Earth's water~\citep{marty06}.  Thus, in the Grand Tack model Earth was delivered water not by C-type asteroids but by the same parent population as for C-type asteroids.  

There remain some issues with the Grand Tack model.  The accretion timescales are much faster for all of the planets than what was typically found in previous models.  This is a consequence of the removal of embryos beyond 1 AU, where growth timescales are long.  In simulations Mars analogs typically form in less than 10 Myr~\citep{obrien13}.  Earth analogs form in 10-20 Myr, with giant embryo-embryo impacts occurring after 20 Myr in only a modest fraction ($\sim20\%$) of simulations.  This is roughly a factor of two faster than the Hf-W constraints~\citep{touboul07,kleine09,konig11}.  However, new simulations show that the accretion timescale of the terrestrial planets can be lengthened to match observations simply by increasing the total embryo-to-planetesimal mass ratio in the annulus, which is itself an unconstrained parameter~\citep{jacobson13}.  


An open question related to the origin of Mercury's small mass is the origin of the {\em inner} edge of the annulus proposed by \cite{hansen09}.  One possibility is that, as embryos grow larger from the inside-out, they also become subject to type 1 migration from the inside-out~\citep{mcneil05,daisaka06,ida08}.  For embryo-mass objects migration is directed inward~\citep{paardekooper11}, so as each embryo forms it migrates inward.  If, by some process, inward-migrating planets are removed from the system (presumably by colliding with the star), then an inner edge in the distribution of {\em surviving} embryos could correspond to the outermost orbital radius at which an embryo formed and was destroyed.  Another possibility is that planetesimals could only form in a narrow annulus.  If a pressure bump~\citep{johansen09} were located in that region it could act to concentrate small particles~\citep{haghighipour03,youdin04} and efficiently form planetesimals (see chapter by Johansen et al.).  


\section{\textbf{DISCUSSION}}

\bigskip
\noindent
\textbf{7.1 Terrestrial planets vs. giant embryos}
\bigskip


We think that Earth formed via successive collisions between planetesimals and planetary embryos, including a protracted stage of giant impacts between embryos.  But does the formation of most terrestrial planets follow the same blueprint as Earth?  

The alternative is that terrestrial exoplanets are essentially giant planetary embryos.  They form from planetesimals or pebbles and do not undergo a phase of giant impacts after the dissipation of the gaseous disk.  This is a wholly reasonable possibility.  Imagine a disk that only forms planetesimals in a few preferred locations, perhaps at pressure bumps.  The planetesimals in each location could be efficiently swept up into a single large embryo, perhaps by the largest planetesimal undergoing a rapid burst of super-runaway pebble accretion.  Isolated giant embryos would evolve with no direct contact with other embryos.  Only if several embryos formed and migrated toward a common location would embryo-embryo interactions become important, and collisions would only occur if a critical number of embryos was present~\citep[the critical number is about 5;][]{morby08,pierens13}.  

Terrestrial planets and giant embryos should differ in terms of their accretion timescales, their atmospheres, and perhaps their geological evolution.  The timescale for the completion of Earth's accretion is at least ten times longer than the typical gas disk lifetime (see \S 2).  Giant embryos must form within the lifetime of the gaseous disk, while the mechanisms to efficiently concentrate are active.  How would Earth be different if it had accreted ten times faster?  The additional heat of formation and from trapped short-lived radionuclides could act to rapidly devolatilize the giant embryo's interior.  However, giant embryos may be able to gravitationally capture thick envelopes of gas from the disk, at least at cooler locations within the disk~\citep{ikoma12}.  The fate of giant embryos' volatiles remain unstudied.  Nonetheless, given that only a very small amount of H and He are needed to significantly inflate a planet's radius~\citep{fortney07}, giant embryos would likely have low bulk densities.  Many low-density planets have indeed been discovered~\citep{marcy13,weiss13}, although we stress that this does not indicate that these are giant embryos.  

How could we tell observationally whether late phases of giant impacts are common?  Perhaps the simplest approach would be to search for signatures of such impacts around stars that no longer harbor gaseous disks.  The evolution of warm dust, detected as excess emission at mid-infrared wavelengths, has recently been measured to decline on 100 Myr timescales~\citep{meyer08,carpenter09,melis10}.  This dust is thought to trace the terrestrial planet-forming region~\citep{kenyon04} and indicates the presence of planetesimals or other large dust-producing bodies in that region.  In some cases the signature of specific minerals in the dust can indicate that it originated in a larger body.  In fact, the signature of a giant impact was reported by \cite{lisse09} around the $\sim$12 Myr-old A star HD 172555.  Given the 1-10 Myr interval between giant impacts in accretion simulations and the short lifetime of dust produced~\citep{kenyon05,melis12}, a direct measure of the frequency of systems in which giant impacts occur will require a large sample of young stars surveyed at mid-infrared wavelengths~\citep[e.g.,][]{kennedy12}.


\bigskip
\noindent
\textbf{7.2 Limitations of the simulations}
\bigskip


Despite marked advances in the last few years, simulations of terrestrial planet formation remain both computationally and physically limited.  Even the best numerical integrators~\citep{chambers99,duncan98,stadel01} can follow the orbits of at most a few thousand particles at $\sim 1$~AU for the $>$100 Myr timescales of terrestrial planet formation.  There are 3-4 orders of magnitude in uncertainty in the sizes of initial planetesimals, and a corresponding 9-12 orders of magnitude uncertainty in the initial number of planetesimals.  It is clear that current simulations cannot fully simulate the conditions of planet formation except in very constrained settings~\citep[e.g.][]{barnes09}.  Simulations thus resort to including planetesimals that are far more massive than they should be.

There exist several processes thought to be important in planet formation that have yet to be adequately modeled.  For example, the full evolution of a planetesimal swarm including growth, dynamical excitation, and collisional grinding has yet to be fully simulated~\citep[but see][]{bromley11,levison12}.  In addition to the numerical and computational challenges, this task is complicated by the fact that the initial distribution and sizes of planetesimals, pebbles and dust remain at best modestly-constrained by models and observations (see chapters by Johansen et al. and Testi et al).  Likewise, the masses, structure and evolution of the dominant, gaseous components of protoplanetary disks is an issue of ongoing study.  


\section{\textbf{SUMMARY}}

This chapter has flown over a broad swath of the landscape of terrestrial planet formation.  We now summarize the take home messages.

1. The term ``terrestrial planet'' is well-defined in the confines of our Solar System but not in extra-solar planetary systems (\S 1).  

2. There exist ample observed and measured constraints on terrestrial planet formation models in the Solar System and in extra-solar planetary systems (\S 2).  

3. There exist two models for the growth of planetary embryos (\S 3).  Oligarchic growth proposes that embryos grow from swarms of planetesimals.  Pebble accretion proposes that they grow directly from cm-sized pebbles.

4. Starting from systems of embryos and planetesimals, the main factors determining the outcome of terrestrial accretion have been determined by simulations (\S 4).  The most important one is the level of eccentricity excitation of the embryo swarm -- by both gravitational self-stirring and perturbations from giant planets -- as this determines the number, masses and feeding zones of terrestrial planets.  

5. The observed systems of hot Super Earths probably formed by either in-situ accretion from massive disks or inward migration of embryos driven by interactions with the gaseous disk (\S 5.1).  The key debate in differentiating between the models is whether rocky planets that accrete in situ could retain thick gaseous envelopes.  

6. The dynamical histories of giant exoplanets are thought to include gas-driven migration and planet-planet scattering.  An inward-migrating gas giant forms low-mass planets interior to strong resonances and stimulates the formation of outer volatile-rich planets.  Dynamical instabilities among giant planets can destroy terrestrial planets or their building blocks, and this naturally produces a correlation between debris disks and terrestrial planets (\S 5.2).  

7. Historical simulations of terrestrial planet formation could not reproduce Mars' small mass (\S 6.1).  This is called the {\em small Mars problem}.  Simulations can reproduce Mars' small mass by invoking large initial eccentricities of Jupiter and Saturn at their current orbital radii.  Invoking early planetesimal-driven migration of Jupiter and Saturn does not produce a small Mars (\S 6.2).  

8. The {\em Grand Tack} model proposes that Jupiter migrated inward to 1.5 AU then back outward due to disk torques before and after Saturn's formation (\S 6.3).  The inner disk was truncated at 1 AU, producing a large Earth/Mars mass ratio.  Water was delivered to the terrestrial planets in the form of C-class bodies scattered inward during the gas giants' outward migration.  

Finally, it remains unclear whether most systems of terrestrial planets undergo phases of giant collisions between embryos during their formation (\S 7.1).

\bigskip
\noindent \textbf{Acknowledgments.} We are grateful to a long list of colleagues whose work contributed to this chapter.  S.N.R. thanks NASA's Astrobiology Institute for their funding via the University of Washington, University of Colorado, and Virtual Planetary Laboratory lead teams.  S.N.R. and A.M. thank the CNRS's Programme Nationale de Planetologie.  KJW acknowledges support from NLSI CLOE.



\end{document}